\documentclass[twocolumn,preprintnumbers,amsmath,amssymb,superscriptaddress]{revtex4-1}

\usepackage{graphicx}

\usepackage{amsmath}
\usepackage{amssymb}
\usepackage{dcolumn}
\usepackage{bm}
\usepackage{upgreek}

\begin{document}
\preprint{}
\title{X-Ray Amplification by Stimulated Brillouin Scattering}

\author{Matthew R. Edwards}
\email{mredward@princeton.edu}
\affiliation{Department of Mechanical and Aerospace Engineering, Princeton University, Princeton, New Jersey, 08544, USA}%
\author{Julia M. Mikhailova}
\email{j.mikhailova@princeton.edu}
\affiliation{Department of Mechanical and Aerospace Engineering, Princeton University, Princeton, New Jersey, 08544, USA}%
\author{Nathaniel J. Fisch}
\email{fisch@princeton.edu}
\affiliation{Department of Astrophysical Sciences, Princeton University, Princeton, New Jersey, 08544, USA}%

\date{\today}

\begin{abstract}
Plasma-based parametric amplification using stimulated Brillouin scattering offers a route to coherent x-ray pulses orders-of-magnitude more intense than those of the brightest available sources. Brillouin amplification permits amplification of shorter wavelengths with lower pump intensities than Raman amplification, which Landau and collisional damping limit in the x-ray regime. Analytic predictions, numerical solutions of the three-wave coupling equations, and particle-in-cell simulations suggest that Brillouin amplification in solid-density plasmas will allow compression of current x-ray free electron laser pulses to sub-femtosecond durations and unprecedented intensities. 
\end{abstract}
\maketitle
Ultrafast sources of high-intensity coherent x rays have dramatically advanced our understanding of physics at atomic spatial and temporal scales \cite{Ackermann2007operation,Emma2010first,Fuchs2015anomalous,Krausz2009attosecond}, permitted high-resolution study of protein, ceramic, and semiconductor structure \cite{Neutze2000potential,Miao2015beyond,Bostedt2016linac}, and, due to the small size of an x-ray diffraction-limited spot, may allow Schwinger-limit intensities \cite{Schwinger1951gauge,Brezin1970pair,Popov1971production} to be reached with moderate pulse energies \cite{Ringwald2001pair,Alkofer2001pair,Popov2001schwinger,Malkin2007compression}. Further advances require amplifying and compressing x-ray pulses beyond the current capabilities of x-ray free-electron lasers (FELs) \cite{Ackermann2007operation,Emma2010first,Fuchs2015anomalous,Bostedt2016linac}, x-ray lasers \cite{Sebban2000full,Zeitoun2004high,Suckewer2009xray,Alessi2011efficient,Rohringer2012atomic}, or high-order-harmonic-generation based sources \cite{Burnett1977harmonic,Mcpherson1987studies,Ferray1988multiple,Gibbon1996harmonic,Lichters1996short,Chang1997generation,Hentschel2001attosecond,Quere2006coherent,Dromey2007bright,Popmintchev2015ultraviolet,Edwards2016waveform}. The parametric amplification of x rays in high-density plasmas using stimulated Raman scattering (SRS) has been suggested for compressing the output of FELs at nanometer wavelengths, potentially reaching attoscond durations \cite{Malkin2007relic,Malkin2007compression,Malkin2009quasitransient,Malkin2010quasitransient,Sadler2015compression}. However, Raman amplification is constrained by Landau and collisional damping at low and high densities and high and low temperatures, respectively. Without the application of extraordinarily large magnetic fields \cite{Shi2017Laser}, the window between these damping mechanisms closes at x-ray wavelengths, rendering SRS amplification impractical for reasonable pump intensities.

We demonstrate that stimulated Brillouin scattering (SBS) is a more promising approach for the amplification of x-ray beams. Even though SBS is less practical than SRS at optical wavelengths, in the x-ray regime the weaker damping of the ion wave is critical. We show that in solid density plasmas with current pump intensities SBS amplification may provide sub-femtosecond x-ray pulses orders-of-magnitude more intense than FEL output.

Consider parametric plasma amplification based on three-wave coupling between a seed beam (frequency $\omega_1$, wavenumber $k_1$), pump beam ($\omega_2$, $k_2$), and Langmuir (SRS) or ion-acoustic (SBS) plasma wave ($\omega_3$, $k_3$). When the resonance conditions for counterpropagating beams are approximately met ($\omega_2 = \omega_3+\omega_1$, $k_2 = k_3-k_1$), energy is efficiently transferred from the pump to the seed beam, resulting in amplification and compression of the seed pulse to up to relativistic intensities and plasma-wave-period durations \cite{Malkin1999}. Since the Langmuir wave has a shorter period than the ion-acoustic wave, Raman amplification tends to give higher growth rates, more intense amplified pulses, and shorter pulse durations \cite{Edwards2016short}, and SRS has been studied in greater depth \cite{Malkin2000,Ping2004,Cheng2005,Ren2008,Yampolsky2008,Ping2009,Vieux2011,Trines2011production,Turnbull2012,Depierreux2014,Toroker2014,Lehmann2014,Edwards2015}. However, because SBS allows the pump and seed to be at almost the same wavelength, it has been considered as an alternative to SRS in both the weakly-coupled regime (WC-SBS), where the plasma mode is the ion acoustic wave \cite{Milroy1977,Andreev1989}, and the strongly-coupled regime (SC-SBS), where the plasma response is a driven ion quasi-mode \cite{Andreev2006,Lancia2010,Lehmann2012,Lehmann2013,Weber2013,Riconda2013,Guillaume2014demonstration,Lehmann2016temperature,Lancia2016signatures,Chiaramello2016role}. For Brillouin amplification it is often necessary to consider both the strongly and weakly-coupled contributions \cite{Edwards2016short,Schluck2016dynamical}, as well as the distinction between SRS and SBS \cite{Jia2016}. Particle-in-cell (PIC) simulations have also suggested that SBS performance improves in a slightly collisional plasma \cite{Humphrey2013effect}. Here we consider the x-ray regime, where SBS may be more practical than SRS because its robustness to damping allows shorter wavelengths, lower pump intensities, and a broader range of plasma densities and temperatures.

An adequate model for x-ray plasma amplification must include treatment of wave damping. Previous works \cite{Malkin2007relic,Malkin2007compression,Malkin2009quasitransient,Malkin2010quasitransient,Balakin2011numerical} have assumed damping of the plasma wave to be dominant, neglecting electromagnetic-wave damping. Although this assumption produces only moderate errors for SRS, the reduced damping of the ion-acoustic mode means that electromagnetic wave damping substantially affects SBS calculations. To find the linear growth rates and estimate the limits of amplification, we use the three-wave coupling equations \cite{Malkin1999}, extending the treatment of Malkin and Fisch for quasitransient backward Raman amplification (QBRA) \cite{Malkin2009quasitransient,Malkin2010quasitransient} to include SBS and collisional damping of the seed. Assuming that an envelope approximation is valid and the phase-matching conditions are met, the fluid model reduces to three-wave coupling equations for the seed ($a_1$), pump ($a_2$), and electron density fluctuation ($n_3$) envelopes (Appendix A). In normalized units (time $\tilde{t} = t\omega_2$, position $\tilde{x} = x \omega_2/c$), the seed and pump equations are:
\begin{align}
\label{eqn:seed}
\left[ \partial_{\tilde{t}} + \tilde{v}_1 \partial_{\tilde{x}} + \tilde{\nu}_1 \right] a_1 &= - \frac{1}{4}  \frac{\omega_2}{\omega_1} N \left[n_3^* a_2\right] \\
\left[ \partial_{\tilde{t}} + \tilde{v}_2 \partial_{\tilde{x}} + \tilde{\nu}_2 \right] a_2 &= \frac{1}{4}   N \left[n_3 a_1\right]
\end{align}
where $a_{1,2} = e A_{1,2} / m_e c$ is the normalized vector potential, $\nu_{1,2} = \tilde{\nu}_{1,2} \omega_2$ is the damping rate, $\tilde{v}_{1,2} = k_{1,2} c / \omega_{1,2} = (1-\omega_{pe}^2/\omega_{1,2}^2)^{1/2}$ is the normalized group velocity for light in terms of the plasma frequency $\omega_{pe} = (4 \pi e^2 n_e/ m_e)^{1/2}$, and $N = n_e/n_c = \omega_{pe}^2/\omega_2^2$ is the plasma number density ($n_e$) normalized by the pump-frequency critical density $n_c = m_e \omega_2^2 / 4\pi e^2$ defined in terms of the electron charge ($e$) and mass $(m_e)$. 

The system is closed with an equation for the electron density, which, for SRS, is
\begin{equation}
\left[ \partial_{\tilde{t}} + \tilde{v}_3 \partial_{\tilde{x}} + \tilde{\nu}_3^R \right] n_3 = - \frac{1}{4}  \frac{1}{\sqrt{N}} \frac{c^2 k_3^2}{\omega_2^2} \left[a_1^* a_2\right]
\end{equation}
where $\tilde{v}_3 = k_3 c/\omega_3$ and $(\omega_3, k_3)$ are found by solving the phase-matching conditions with the dispersion relation $\omega_3 = \omega_{pe}$. For SBS,
\begin{equation}
\label{eqn:densSBS}
 \left[ \frac{i}{2} \frac{\omega_2}{\omega_3} \partial^2_{\tilde{t}}  + \partial_{\tilde{t}} + \tilde{\nu}_3^B \right] n_3 = -\frac{1}{4} \frac{Zm_e}{m_i} \frac{\omega_2}{\omega_3} \frac{c^2 k_3^2}{\omega_2^2} \left[a_1^* a_2\right]
\end{equation}
where $Z$ gives the ion charge state, $m_i$ is the ion mass, and the plasma response ($\omega_3, k_3$) satisfies the dispersion relation \cite{Edwards2016short}:
\begin{equation}
\left[\omega_3^2 - c_s^2 k_3^2\right]\left[\omega_3^2 - 2\omega_2 \omega_3 -c^2 k_3^2 + 2c^2 k_2 k_3\right] = \frac{k_3^2 c^2 a_{0}^2  \omega_{pe}^2}{4 m_i/Zm_e}
\end{equation}
with $c_s = \sqrt{ZT_e/m_i}$ for electron temperature $T_e \gg T_i$ \cite{Nicholson1983}. In Eq.~(\ref{eqn:densSBS}), the group velocity term for SBS is small and has been neglected. The first derivative in time is dominant for WC-SBS, where $\omega_3 = c_s k_3$, and the second derivative dominates for SC-SBS.

Assuming negligible pump depletion, the asymptotic linear growth rate of the seed may be calculated from the impulse response of the seed-plasma wave system \cite{Bobroff1967impulse}, as previously applied for QBRA \cite{Malkin2009quasitransient,Malkin2010quasitransient}, or by calculating a self-similar solution for the system as $\tilde{t} \to \infty$ (derived in Appendix B). Including collisional damping of the seed, the growth rate becomes
\begin{equation}
\hat{\Gamma}_{R,B_{wc}} = \frac{1}{2} \left[4 \tilde{\Gamma}_{R,B_{wc}}^2 + \left(\tilde{\nu}_3^{R,B} - \tilde{\nu}_1 \right)^2\right]^{\frac{1}{2}} - \frac{\tilde{\nu}_3^{R,B} + \tilde{\nu}_1}{2}
\label{eqn:SRSgrowth}
\end{equation}
for SRS and WC-SBS, where $\tilde{\nu}_1 = N\tilde{\nu}_{ei}\omega_2/\omega_1$ \cite{Ginzburg1970}, and $\tilde{\nu}_3^R = \tilde{\nu}_{\textrm{Lnd}} +\tilde{\nu}_{\textrm{ei}}/4$ \cite{Malkin2010quasitransient}. For SBS with $T_i \ll T_e$, previous work \cite{Epperlein1992Damping} suggests a damping coefficient $\tilde{\nu}_3^B = 0.01 \times 2c_s/c$ for conditions considered here; this rate is small, and even substantial variations do not affect the final results. For SC-SBS and SBS in the intermediate regime, the effective growth rate $\hat{\Gamma}_B = \kappa$ is the largest real root of the cubic equation:
\begin{equation}
C_{sc} \kappa^3 +\left(C_{sc} \tilde{\nu}_1 + C_{wc}\right) \kappa^2 + \left(C_{wc} \tilde{\nu}_1 + \tilde{\nu}_3^B\right) \kappa + K = 0
\label{eqn:SBSgrowth}
\end{equation}
where $C_{sc} = i\omega_2 / 2\omega_3$, $C_{wc} = 1$ (assumed 0 for SC-SBS), and $K = \tilde{\nu}_1\tilde{\nu}_3^B  - N Z m_e c^2 k_3^2 / 16 \omega_1 \omega_3 m_i$. The normalized collisional and Landau damping rates are \cite{Malkin2010quasitransient}:
\begin{equation}
\tilde{\nu}_{\textrm{ei}} = \frac{2 \sqrt{2}}{3 \sqrt{N\pi}} \frac{r_e \Lambda_C \omega_2}{c q_T^{3/2}} \qquad \tilde{\nu}_{\textrm{Lnd}} = \frac{\sqrt{N \pi}}{(2q_T)^{3/2}} e^{-\frac{1}{q_T}}
\end{equation}
where $q_T = 4T_e/Nm_ec^2$, $\Lambda_C = \ln 12 \pi \lambda_D^3 n_e /Z$ is the Coulomb logarithm, $\lambda_D = (T_e/4\pi n_e e^2)^{1/2}$ is the Debye length, and $r_e = e^2/m_ec^2$. The undamped Raman growth rate, with $\omega_3 = \omega_{pe}$, is \cite{Kruer2003}:
\begin{equation}
\tilde{\Gamma}_R = \frac{a_0 k_3 c N^{\frac{1}{4}}}{4\sqrt{\omega_1\omega_2}}
\end{equation}
where $a_0$ is the pump field strength ($a_2$) before interaction. The Brillouin growth rate may be evaluated numerically in the intermediate regime \cite{Edwards2016short}, with the weak and strong coupling limits \cite{Kruer2003}:
\begin{equation}
\tilde{\Gamma}_{B_{wc}} = \frac{a_0 c}{2\sqrt{2}} \left[\frac{Z m_e N k_2}{m_i \omega_2 c_s}\right]^{\frac{1}{2}} \quad \tilde{\Gamma}_{B_{sc}} = \frac{\sqrt{3}}{2}\left[\frac{a_0^2 c^2 k_2^2 N}{2 \omega_2^2 m_i/ Z m_e}\right]^{\frac{1}{3}}
\end{equation}

\begin{figure}
\centering
\includegraphics[width=\linewidth]{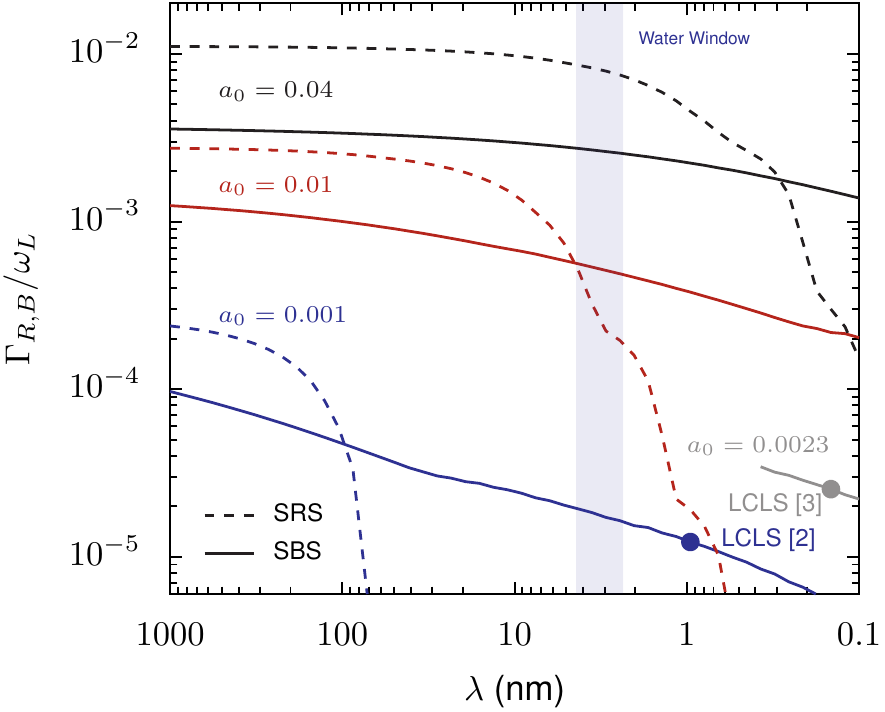}
\caption{The maximum linear growth rate ($\Gamma_{R,B}$), normalized by the pump frequency ($\omega_L=\omega_2$), for SRS (dashed lines) or SBS (solid lines) against the pump wavelength ($\lambda = 2\pi c/\omega_L$) at varied pump strength ($a_0$). Density and temperature are chosen at each wavelength to maximize the growth rate. The water window between the K-absorption edges of oxygen (2.34 nm) and carbon (4.4 nm) is important for biological imaging. Growth rates corresponding to approximate pump strengths and wavelengths demonstrated at the Linac Coherent Light Source (LCLS) \cite{Emma2010first,Fuchs2015anomalous} are marked.}
\label{fig1}
\end{figure}

The growth rate ($\hat{\Gamma}_{R,B}$) of the seed pulse in the linear (constant pump amplitude) regime is a simple metric for comparing SRS and SBS. Apart from dictating the plasma length required to amplify an initially small seed, the linear growth rate determines whether the amplification process out-competes deleterious effects like filamentation, dispersion, and forward scattering. Using Eqs.~(\ref{eqn:SRSgrowth}) and (\ref{eqn:SBSgrowth}), the maximum SRS and SBS growth rates achievable over all densities and electron temperatures at a particular pump wavelength are plotted in Fig.~\ref{fig1}. At optical frequencies SRS has a higher growth rate than SBS, but collisional and Landau damping create a pump-amplitude-dependent cutoff in the x-ray regime, where $\hat{\Gamma}_R$ rapidly drops; for example, below $\lambda = 4$ at $a_0 = 0.01$, $\hat{\Gamma}_B > \hat{\Gamma}_R$. Since $a_0$ scales with wavelength, the intensities required to achieve a sufficient SRS growth rate below $\lambda = 1$ nm are currently impractical, e.g. at $\lambda = 1$ \AA, $a_0 = 0.05$ corresponds to $3\times 10^{23}$ W/cm$^2$. Note that all SBS results presented in this work assume a full-ionized proton plasma ($Z = 1$, $m_i = 1836m_e$) with negligible ion temperature; amplifier performance will decrease as $Z/m_i$ decreases. 

\begin{figure}
\centering
\includegraphics[width=\linewidth]{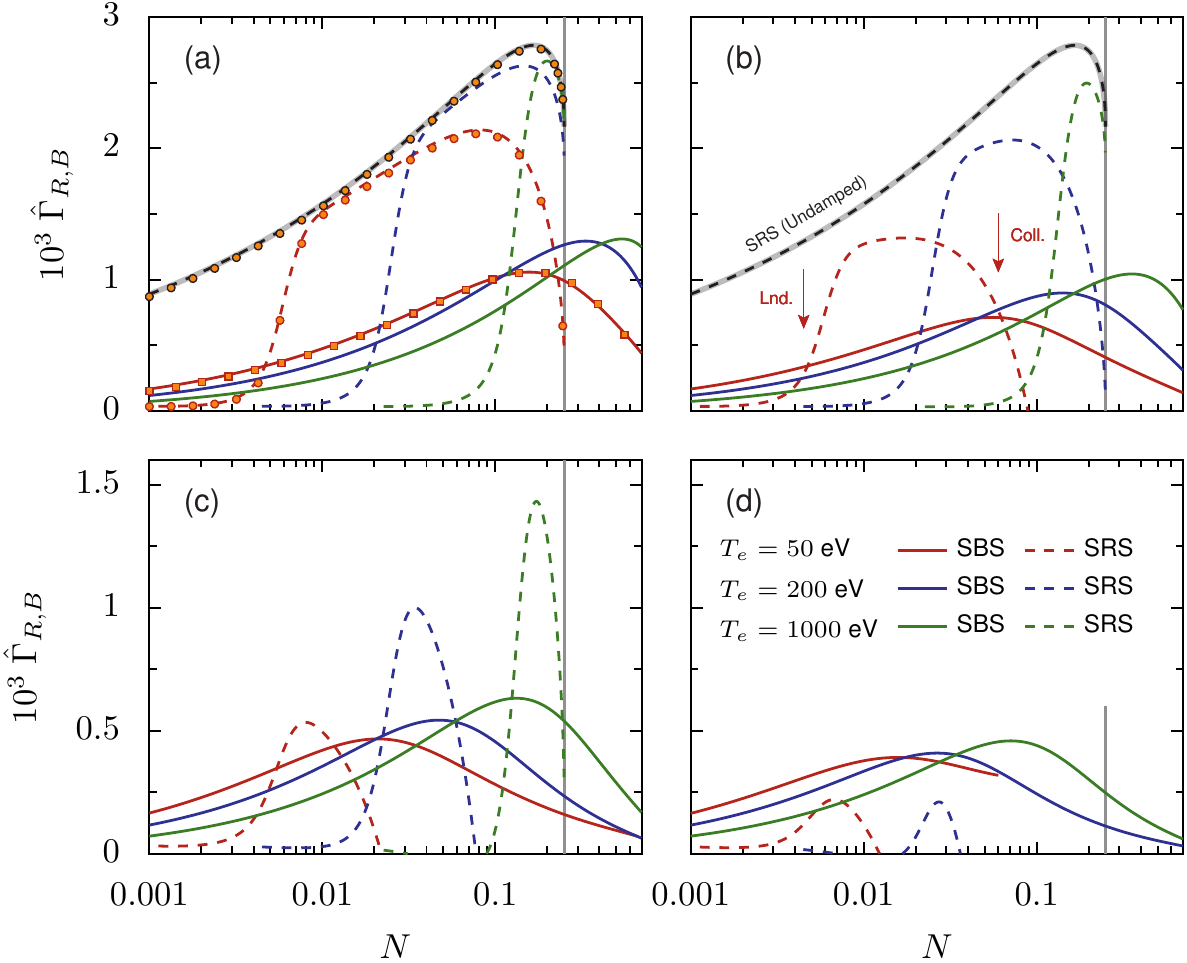}
\caption{The effective linear growth rate of SRS and SBS for $a_0 = 0.01$ and varied $\lambda$, $T_e$, and $N$. (a) $\lambda = 1000$ nm, (b) $\lambda = 100$ nm, (c) $\lambda = 10$ nm, and (d) $\lambda = 3$ nm. The gray vertical line indicates the upper limit of valid densities for SRS ($N = 0.25$). The orange points are from numerical solution of the wave-coupling equations.}
\label{fig2}
\end{figure}

The plasma conditions optimal for SRS and SBS differ, with SBS generally more flexible and favoring higher temperatures. Fig.~\ref{fig2} shows the damped SRS and SBS growth rates [from Eqs.~(\ref{eqn:SRSgrowth}) and (\ref{eqn:SBSgrowth})] at $a_0 = 0.01$. Increased collisional damping reduces the growth rate at high densities, and Landau damping of the Langmuir wave sets the low-density cutoff for SRS. To numerically solve the wave-coupling equations [Eqs.~(\ref{eqn:seed})-(\ref{eqn:densSBS})], we employ a step-shifting algorithm for the spatial derivatives \cite{ClarkThesis,Clark2003operating,Clark2003particle} and a fourth-order Runge-Kutta scheme for the time derivatives; numerical solutions are marked with points. 

\begin{figure}
\centering
\includegraphics[width=\linewidth]{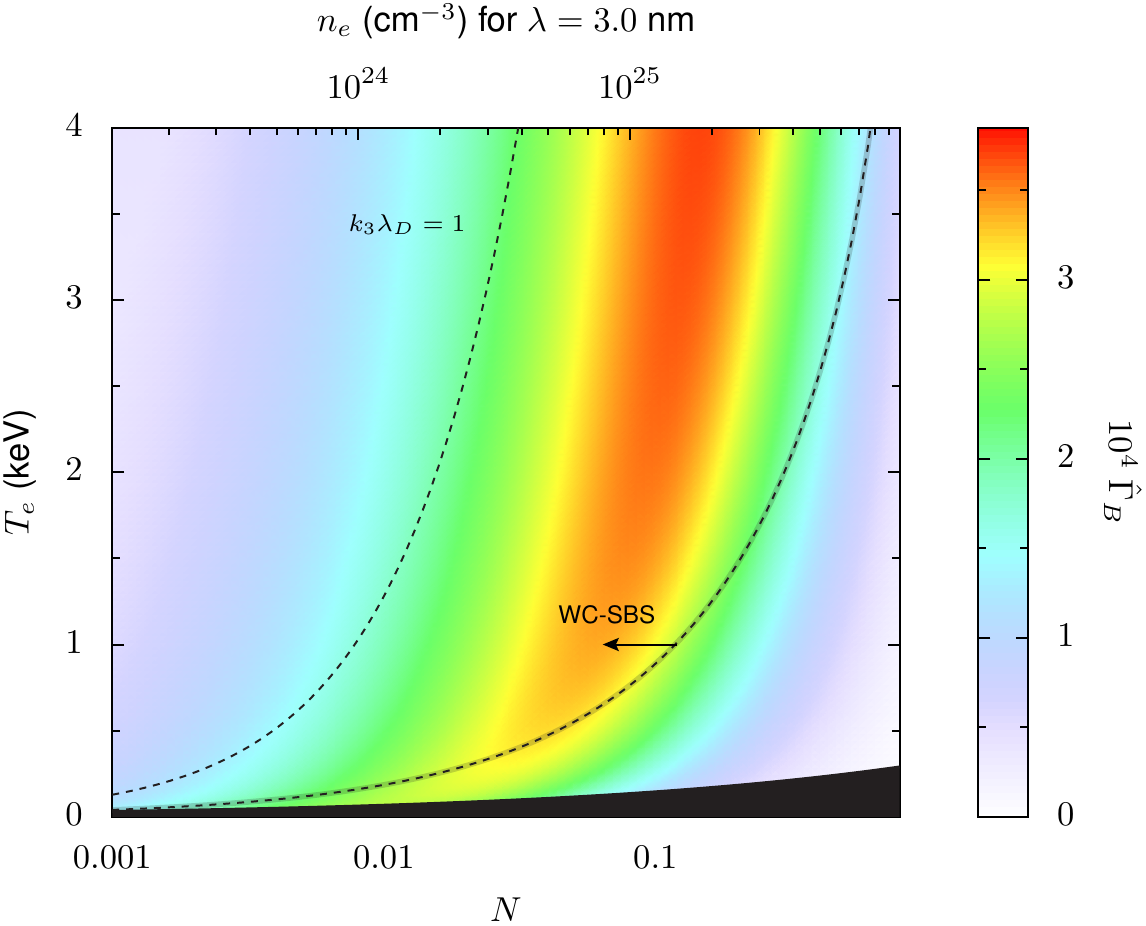}
\caption{The growth rate of SBS against density ($N$) and temperature ($T_e$) for pump wavelength $\lambda = 3$ nm and $a_0 = 0.01$. The threshold $\Lambda = 0.25 = (Zm_e/m_i)(\omega_{pe} a_0 c)^2 / (16 \omega_2 k_2 c_s^3)$ \cite{Edwards2016short} between WC-SBS and SC-SBS is marked by a dashed line. The black region at bottom excludes parameters where the Coulomb-logarithm collisional model is substantially invalid ($\Lambda_C < 1$).}
\label{fig3}
\end{figure}

For optical wavelengths SC-SBS is usually considered more promising than WC-SBS, due to a higher growth rate and shorter permitted pulse durations \cite{Andreev2006}. However, in extending SBS to the x-ray regime, we find that heavier collisional damping at the high densities and low temperatures associated with SC-SBS means the highest growth rates may be in the weakly-coupled regime for reachable pump intensities, (Fig.~\ref{fig3}). Filamentation, the modulational instability, forward Raman scattering, dispersion, and collisional damping of the pump are more significant at higher plasma densities, so the highest output seed intensities will be reached for lower plasma densities than those corresponding to the highest growth rates. 

\begin{figure}
\centering
\includegraphics[width=\linewidth]{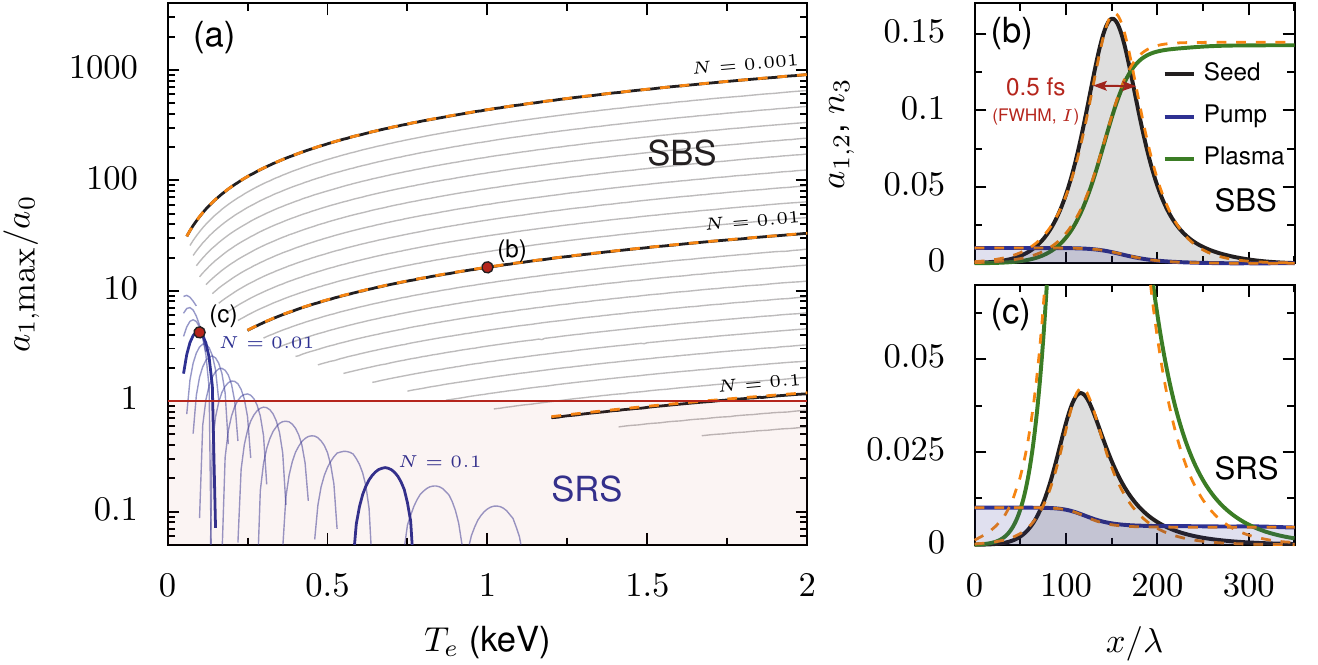}
\caption{(a) Maximum seed-to-pump amplification factor (asymptotic limit) produced by SRS or SBS at $\lambda = 3$ nm and $a_0 = 0.01$. The orange dashed lines mark the analytically calculated values for SBS, neglecting $\tilde{\nu}_3$. Thin lines are logarithmically spaced in density, with $N = 0.001, 0.01, 0.1$ marked by bold lines. (b,c) The wave envelope limits from solution of the three-wave equations for selected SBS (b) and SRS (c) pulses at $N = 0.01$, $T_e = 1$ keV and $ N = 0.01$, $T_e = 100$, respectively. In (b,c) the orange dashed lines mark solutions to Eqs.~(\ref{eqn:ss1})-(\ref{eqn:ss2}).}
\label{fig4}
\end{figure}

Without seed damping, the wave-coupling equations predict infinite amplification (until the underlying assumptions fail); with $\tilde{\nu}_3 \ne 0$ the solution approaches a limit at infinite time. This asymptote constrains the maximum possible amplification, since any seed pulse will converge towards it, i.e. a larger seed pulse will decrease in amplitude. In this limit Eqs.~(\ref{eqn:seed})-(\ref{eqn:densSBS}) reduce (Appendix C) to:
\begin{align}
\tilde{\nu}_1 a_1 &= K_1 n_3 a_2 \label{eqn:ss1} \\
(\tilde{v}_2 -\tilde{v}_1) \partial_{\tilde{x}} a_2 &= K_2 n_3 a_1 \\
-\tilde{v}_1  \partial_{\tilde{x}} n_3 + \tilde{\nu}_3 n_3 &= K_3 a_1 a_2 \label{eqn:ss2}
\end{align}
where $K_1 = -\omega_2 N / 4 \omega_1$, $K_2 = N/4$, $K_3^R = - c^2 k_3^2 / 4 \omega_2^2 \sqrt{N}$, and $K_3^B = Z m_e c^2 k_3^2 / 4 m_i \omega_3 \omega_2$ for SBS. An analytic solution of these equations is possible if $\tilde{\nu}_3 \ll \tilde{\nu}_1$, as is the case for SBS. This yields
\begin{equation}
a_1(\tilde{x}, \tilde{t}\to \infty) = \frac{K_1 a_{0}^2}{\tilde{\nu}_1 \sqrt{2 D_1}} \frac{e^{D_2 a_{0}^2\tilde{x}}}{1 + (1/2D_2) e^{2D_2 a_{0}^2\tilde{x}}}
\end{equation}
where $D_1 = K_1 K_2 / \tilde{\nu}_1 (\tilde{v}_2 - \tilde{v}_1)$ and $D_2 = -K_1K_3 / \tilde{v}_1 \tilde{\nu}_1$, with maximum value:
\begin{equation}
a_{1, \textrm{max}} = \frac{K_1 a_{0}^2}{2 \tilde{\nu}_1} \left[\frac{K_3}{K_2} \left(1 + \frac{\tilde{v}_2}{\tilde{v}_1}\right) \right]^{\frac{1}{2}}
\end{equation}
The maximum amplification factor $(a_{1, \textrm{max}}/a_0)$ at $a_0 = 0.01$ and $\lambda_0 = 3$ nm as calculated from these equations is shown in Fig.~\ref{fig4}. For SRS, where $\tilde{\nu}_3$ cannot be neglected, the solution is found numerically. For these parameters the amplification factors achievable with SBS are far higher than with SRS and the amplified pulse in Fig.~\ref{fig4}b has an intensity full-width-half-maximum of 0.5 fs, suggesting that at its limit SBS may produce intense sub-femtosecond x-ray pulses. Higher amplification factors are associated with lower densities, which give lower growth rates, so a practical device might use a high-to-low density gradient to combine high initial growth rates with large final amplitude limits. Relativistic and dispersive effects will stop growth before the indicated asymptotes. We can estimate the effect of the modulational instability by calculating the energy transferred from pump to seed in the pump-depletion regime during a normalized modulational growth time $2/N a_1^2$ for the final seed intensity \cite{Malkin1999}. Using a compressed duration $200$ $\lambda/c$ at $N=0.01$ and assuming that half the pump energy is transferred to the seed, we find that  $a_0=0.01$ can produce $a_1 = 0.1$ in a modulational time, or a two order of magnitude increase in intensity, which is similar to the limits suggested by the collisional asymptote.

\begin{figure}
\centering
\includegraphics[width=\linewidth]{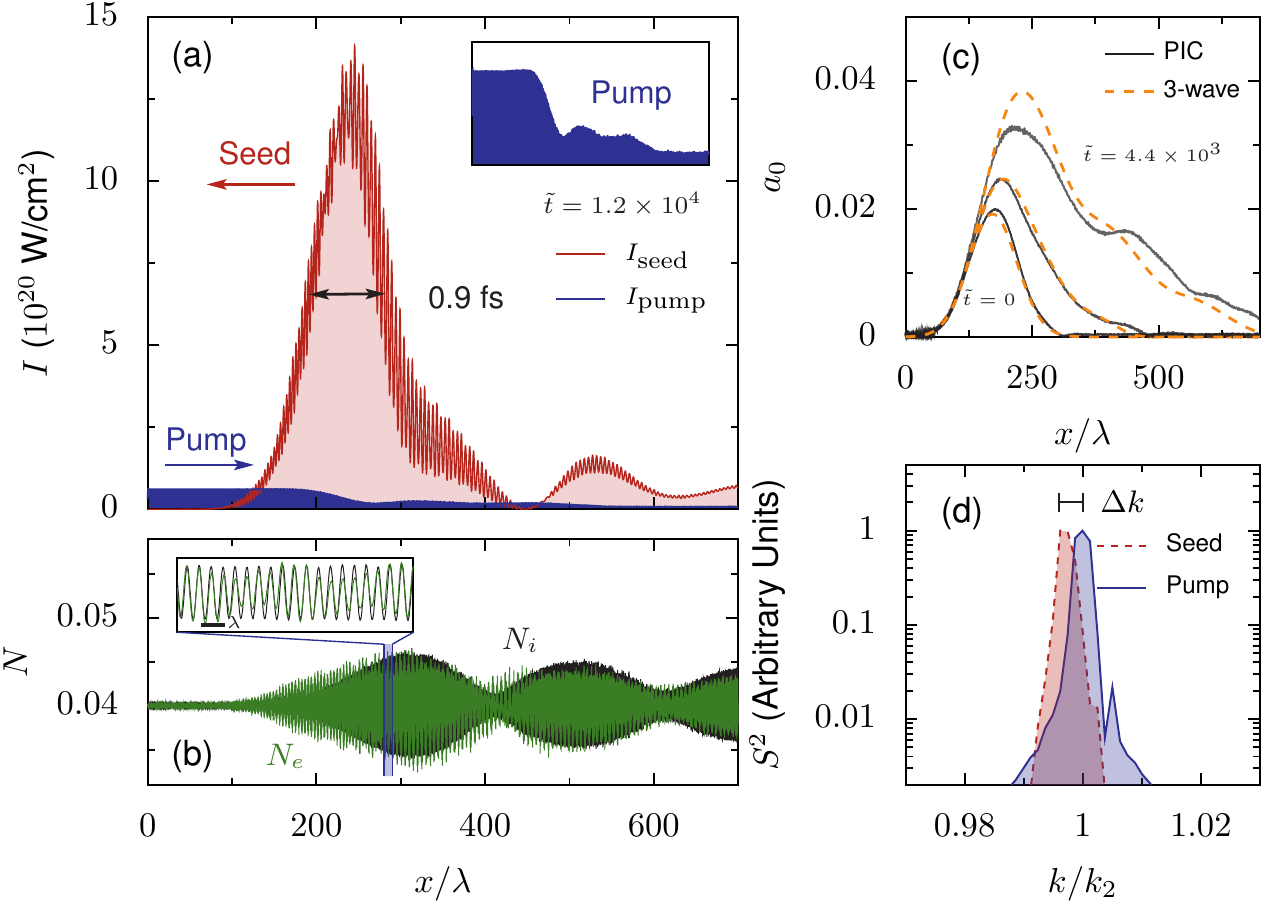}
\caption{PIC simulation (EPOCH \cite{Arber2015contemporary}) of Brillouin amplification with $\lambda = 3$ nm, $a_0 = 0.02$, $N = 0.04$, $T_e = 2$ keV, $T_i = 10$ eV, and $m_i = 1836 m_e$. (a) Intensity ($I$) of seed (red) and pump (blue) beams at $\tilde{t} = 1.2\times 10^4$ (19 fs) from PIC calculations. Inset: Pump intensity only, showing $85$\% depletion. (b) Electron ($N_e$, green) and ion ($N_i$, black) density. Inset: a $10\lambda$ interval of (b), showing $0.5 \lambda$ period of plasma density fluctuations. (c) Evolution of seed envelope from initial time ($\tilde{t} = 0$) to $\tilde{t} = 4.4\times 10^3$ for PIC (solid gray) and three-wave (dashed orange) calculations. (d) Spectra of pump and seed at $\tilde{t} = 8.2\times10^3$. $\Delta k$ is the wavenumber downshift required for the SBS phase matching condition and agrees with the observed shift. Here $\nu_{ei}^{-1} \approx 1$ fs, and collisions have a small effect on the interaction. The PIC calculations use 80 cells/$\lambda$, 1000 particles per cell, and a simulation window moving at the seed group velocity. No amplification is observed in an equivalent simulation with immobile ions.}
\label{fig5}
\end{figure}

The above analysis is limited by the assumptions that the fluid model is valid (i.e. kinetic effects are neglected), that damping is adequately described by a linear model, and that density fluctuations may be considered small. These assumptions can be checked by PIC simulations, which include kinetic, dispersive, and relativistic effects. Using the code \verb+EPOCH+ \cite{Arber2015contemporary}, we demonstrate in Fig.~\ref{fig5} that under these conditions the three-wave model makes reasonable predictions for solutions of the full system; the $\lambda = 3$ nm pump amplifies the seed pulse to 20 times the pump intensity in 19 fs, producing a 0.9-fs-duration output pulse. The density fluctuations (Fig.~\ref{fig5}b) and amplified spectra (Fig.~\ref{fig5}d) are characteristic of Brillouin amplification, and the seed envelope (Fig.~\ref{fig5}c) evolves in agreement with the three-wave model. The slightly lower growth in the PIC calculation and the appearance of envelope modulations result from forward Raman scattering, which is overestimated due to the particle discreteness in PIC simulations. 

Consider a focused intensity of $10^{19}$ W/cm$^2$ at $\lambda = 3$ nm, which lies near the capabilities of current and proposed advanced FEL sources \cite{Ackermann2007operation,Emma2010first,Fuchs2015anomalous}. In a plasma with density near $5 \times 10^{23}$ cm$^{-3}$ and temperature 500 eV, slightly higher than compressed solid hydrogen \cite{Weir1996metallization} and lower than fully-ionized aluminum, the SBS growth rate (in hydrogen) will be around 100 ps$^{-1}$. Metal targets will allow higher electron densities than hydrogen, but the lower $Z/m_i$ ratio will reduce the amplification growth rate, and the higher collision frequency will cause additional damping. The modulational and collisional limits, as described previously, give for these conditions a maximum output pulse intensity 100-1000 times greater than the pump intensity, up to $10^{21}-10^{22}$ W/cm$^2$. With a seed provided by laser-driven high-order harmonic generation, SBS could be used to amplify and compress the output of large-scale x-ray FEL by several orders of magnitude in a millimeter scale plasma. For comparison, at this wavelength and pump intensity, SRS is too heavily damped to produce any significant amplification for any plasma density or temperature.

\begin{figure}
\centering
\includegraphics[width=\linewidth]{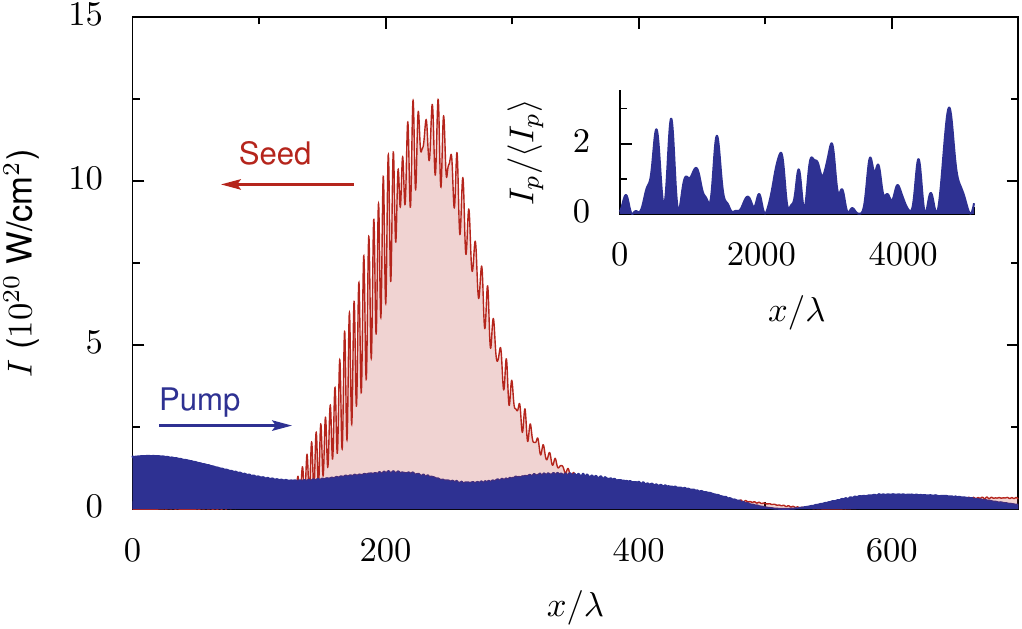}
\caption{PIC simulation (EPOCH \cite{Arber2015contemporary}) of Brillouin amplification in an incoherent pump (bandwidth FWHM $0.4$\% of $\omega_0$) with $\lambda = 3$ nm, $a_0 = 0.02$, $N = 0.04$, $T_e = 2$ keV, $T_i = 10$ eV, and $m_i = 1836 m_e$ at $\tilde{t} = 6.3 \times 10^3$. Apart from pump coherence, conditions and simulation parameters are the same as those in Fig.~\ref{fig5}. Inset: pump intensity envelope (before interaction) showing amplitude modulations due to incoherence.}
\label{fig6}
\end{figure}

Although FEL beams are only quasi-coherent, previous studies have suggested that parametric plasma amplification is relatively resilient to moderate amplitude and phase fluctuations \cite{Balakin2003noise,Solodov2003,Fraiman2002,Kirkwood2011observation}. Current x-ray FELs have bandwidths on the order of 0.1-1\% of $\omega_0$ \cite{Geloni2010coherence,Vartanyants2011coherence}. In Fig.~\ref{fig6}, a PIC simulation of amplification of x-ray Brillouin amplification using a pump bandwidth $\Delta \omega = 0.4\% \times \omega_0$ (FWHM) with corresponding coherence time around 0.5 fs shows that the inclusion of finite coherence does not impede the interaction, provided that the coherence time is not short compared to the final amplified pulse. The pump is constructed by randomly choosing the phase and frequency of 1000 components from a Gaussian frequency distribution, producing the modulated intensity envelope shown as an inset in Fig.~\ref{fig6}. The amplified pulse shown in Fig.~\ref{fig6} has an intensity and duration similar to that found using a coherent pump under the same conditions (Fig.~\ref{fig5}). It should be noted that the seed pulse maintains its coherence, so SBS may be useful for cleaning FEL output to produce high intensity coherent pulses from a lower-intensity quasi-coherent source. 

To conclude, this work predicts an amplification regime where stimulated Brillouin scattering is significantly more useful than Raman scattering, allowing coherent x-ray pulses with unprecedented intensities. In this regime, plasma-based x-ray amplification may be possible with available FEL beamlines.

\begin{acknowledgments}
This work was supported by NNSA Grant No.~DENA0002948, AFOSR Grant No.~FA9550-15-1-0391, and NSF Grant No.~PHY 1506372. M.R.E.~ acknowledges the support of the NSF. Computing support for this work came from the High Performance Computing Center at Princeton University and the Lawrence Livermore National Laboratory (LLNL) Institutional Grand Challenge program. The EPOCH code was developed as part of the UK EPSRC 300 360 funded project EP/G054940/1. We would like to thank Q.~Jia, K.~Qu, Y.~Ping, S.~Meuren, and R.~Berger for useful discussions.   
\end{acknowledgments}

\appendix

\section{Derivation of the Three-Wave Coupling Equations}
The three-wave coupling equations play a crucial role in the analysis of Raman and Brillouin amplification. Since the work presented here relies on a more general treatment of the governing equations than is usually necessary, we provide a derivation for reference below. 

We start with the Maxwell-fluid equations, which describe the coupling between light and plasma based on a fluid model of the electron and ion motion \cite{Kruer2003}. This model does not include kinetic and non-continuum effects from first principles, though Landau and collisional damping are incorporated empirically through added terms, so the model is only applicable where other kinetic effects are unimportant and the expressions for the damping terms are valid. Considering a seed beam (vector potential $\mathbf{A}_1$), a pump beam (vector potential $\mathbf{A}_2$) and a plasma of number density $n_{e,0}$, the governing equations are:
\begin{equation}
\label{eqn:A1}
\left[ \partial_t^2 - c^2 \partial_x^2 +\omega_{pe}^2 + \nu_{1} \partial_t \right] \mathbf{A}_1= -\omega_{pe}^2 \frac{n_{e,1}}{n_{e,0}} \mathbf{A}_2
\end{equation}
\begin{equation}
\left[ \partial_t^2 - c^2 \partial_x^2 +\omega_{pe}^2 + \nu_{2} \partial_t \right] \mathbf{A}_2= -\omega_{pe}^2 \frac{n_{e,1}}{n_{e,0}} \mathbf{A}_1
\end{equation}
where $\nu_{1,2}$ are damping coefficients for the light wave, $\omega_{pe} = \sqrt{4\pi e^2 n_{e,0}/m_e}$ is the plasma frequency, and $n_{e,1}$ is the amplitude of small electron number density fluctuations. To close the system, we write an equation for the coupling of an electron plasma wave (SRS),
\begin{equation}
\left[ \partial_t^2 - 3 v_e^2 \partial_x^2 +\omega_{pe}^2 + \nu_{3}^R \partial_t \right] \frac{n_{e,1}}{n_{e,0}} = \frac{e^2}{m_e^2 c^2} \partial_x^2 \left(\mathbf{A}_1 \cdot \mathbf{A}_2\right)
\end{equation}
or an ion-driven electron density perturbation (SBS),
\begin{equation}
\left[ \partial_t^2 - c_s^2 \partial_x^2 + \nu_{3}^B \partial_t \right] \frac{n_{e,1}}{n_{e,0}} = \frac{Z e^2}{m_e m_i c^2} \partial_x^2 \left(\mathbf{A}_1 \cdot \mathbf{A}_2\right)
\end{equation}
to the laser fields \cite{Kruer2003}. Here, one ion species is assumed, with $Z$ as the ion charge and $m_i$ as the ion mass. The plasma sound speed is $c_s = \sqrt{ZT_e/m_i}$, assuming that $T_i \ll T_e$ and that the fluctuations are quasi-neutral \cite{Nicholson1983}.   

Assuming that the laser fields may be described as an envelope over a high frequency carrier wave, for linear polarization, the vector potential is written:
\begin{align}
\mathbf{A}_{1,2}\left(x,t\right) =\: & \frac{m_e c^2}{2 e} a_{1,2}\left(x,t\right) e^{i\left(k_{1,2}x - \omega_{1,2} t \right)} \mathbf{\hat{y}} \nonumber \\ 
&+\frac{m_e c^2}{2 e} a_{1,2}^* \left(x,t\right) e^{-i\left(k_{1,2}x - \omega_{1,2} t \right)} \mathbf{\hat{y}}
\end{align}
where $k_1$ and $k_2$ contain the directionality of the wave propagation, i.e. $k_1 < 0$ corresponds to propagation in the negative $x$ direction. The second term is the complex conjugate (c.c.) of the first. For this article, we consider only linear polarization. 

Similarly, the density fluctuations may be written:
\begin{equation}
\frac{n_{e,1}(x,t)}{n_{e,0}}  = \frac{i}{2} n_3(x,t) e^{i\left(k_{3}x - \omega_{3} t \right)} + \textrm{c.c.}
\end{equation}
where we have notated $n_3$ with the subscript $3$ as a reminder that it represents the third fluctuating quantity. We assume that the phase matching conditions are met: $k_3 = k_2 - k_1$ and $\omega_3 = \omega_2 - \omega_1$. Substituting these definitions (for linear polarization) into Eq.~(\ref{eqn:A1}), evaluating derivatives of the high frequency components, and dropping non-resonant terms, we have:
\begin{widetext}
\begin{equation}
\label{eqn:A8}
\left[ \partial_t^2 - 2i \omega_1 \partial_t - \omega_1^2 -c^2 \partial_x^2 + 2 i c^2 k_1 \partial_x + c^2 k_1^2 +\omega_{pe}^2 +\nu_1 \partial_t - i\nu_1 \omega_1 \right] a_1 = \frac{-i \omega_{pe}^2}{2} n_3 a_2 
\end{equation}
with a similar expression for $a_2$. We assume that the light propagation satisfies the dispersion relation \cite{Ginzburg1970}:
\begin{equation}
c^2 k_1^2 = \omega_1^2\left[1 - \frac{\omega_{pe}^2}{\omega \left(\omega-i\nu_{ei}\right)} \right] \approx \omega_1^2 \left[ 1- \frac{\omega_{pe}^2}{\omega_1^2} \left( 1 + \frac{i \nu_{ei}}{\omega_1}\right) \right] = \omega_1^2 - \omega_{pe}^2 - i  \nu_{ei} \omega_1 \frac{\omega_{pe}^2}{\omega_1^2}
\end{equation}
\end{widetext}
Substituting this dispersion relation into Eq.~(\ref{eqn:A8}), with $\nu_1 = \nu_{ei} \omega_{pe}^2/\omega_1^2$ and noting that $\partial_t \ll \omega_1$ and $\partial_x \ll k_1$, we have:
\begin{equation}
\left[ \omega_1 \partial_t + c^2 k_1 \partial_x + \omega_1 \nu_1 \right] = \frac{1}{4} \frac{\omega_{pe}^2}{2} n_3^*a_2
\end{equation}
To simplify this equation, we use the normalized variables $\tilde{t} = t\omega_2$, $\tilde{x} = x \omega_2/c$, and $\tilde{\nu}_1 = \nu_1/\omega_2$, and write the normalized group velocity as $\tilde{v}_1 = c k_1/\omega_1$. We then have:
\begin{equation}
\left[ \partial_{\tilde{t}} + \tilde{v}_1 \partial_{\tilde{x}} + \tilde{\nu}_1 \right] a_1 = - \frac{1}{4}  \frac{\omega_2}{\omega_1} N \left(n_3^* a_2\right)
\end{equation}
where $N = \omega_{pe}^2/\omega_2^2 = n_e/n_c$. The pump equation similarly simplifies to:
\begin{equation}
\left[ \partial_{\tilde{t}} + \tilde{v}_2 \partial_{\tilde{x}} + \tilde{\nu}_2 \right] a_2 = \frac{1}{4}   N \left(n_3 a_1\right)
\end{equation}

Using the same approach, the density equations for SRS and SBS can be written in wave-coupling form. For the Langmuir wave (SRS):
\begin{equation}
\left[ \partial_{\tilde{t}} + v_3 \partial_{\tilde{x}} + \tilde{\nu}_3^R \right] n_3 = - \frac{1}{4}  \frac{1}{\sqrt{N}} \frac{c^2 k_3^2}{\omega_2^2} \left(a_1^* a_2\right)
\end{equation}
For SBS, we cannot in general make the assumption $\partial_t \ll \omega_3$ and are therefore left with both first and second order time derivatives on the left-hand side. However, the group velocity $\tilde{v}_3$ is in this case negligible, so all spatial derivatives may be dropped. This yields the following expression for the SBS plasma response:
\begin{equation}
 \left[ \frac{i}{2} \frac{\omega_2}{\omega_3} \partial^2_{\tilde{t}}  + \partial_{\tilde{t}} + \tilde{\nu}_3^B \right] n_3 = -\frac{1}{4} \frac{Zm_e}{m_i} \frac{\omega_2}{\omega_3} \frac{c^2 k_3^2}{\omega_2^2} \left(a_1^* a_2\right)
\end{equation}
which reduces to the strongly-coupled and weakly-coupled limits, as described in Ref.~\cite{Edwards2016short}, if the second-order or first-order time derivatives are neglected, respectively. 

The full set of equations describing wave coupling for Raman and Brillouin scattering is given below, where Eqs.~(\ref{eqn:Aseed}) and (\ref{eqn:Apump}) are used for all models, Eq.~(\ref{eqn:SRS}) for SRS, and Eqs.~(\ref{eqn:SBS1})-(\ref{eqn:SBS3}) for the different forms of SBS. 
\begin{widetext}
\begin{align}
\textrm{(Seed)} \qquad  \qquad& \left[ \partial_{\tilde{t}} + v_1 \partial_{\tilde{x}} + \tilde{\nu}_1 \right] a_1 = - \frac{1}{4}  \frac{\omega_2}{\omega_1} N \left(n_3^* a_2\right) \label{eqn:Aseed} \\ \nonumber\\
\textrm{(Pump)} \qquad  \qquad& \left[ \partial_{\tilde{t}} + v_2 \partial_{\tilde{x}} + \tilde{\nu}_2 \right] a_2 = \frac{1}{4}   N \left(n_3 a_1\right) \label{eqn:Apump} \\ \nonumber\\
\textrm{(SRS)} \qquad  \qquad& \left[ \partial_{\tilde{t}} + v_3 \partial_{\tilde{x}} + \tilde{\nu}_3^R \right] n_3 = - \frac{1}{4}  \frac{1}{\sqrt{N}} \frac{c^2 k_3^2}{\omega_2^2} \left(a_1^* a_2\right)\label{eqn:SRS}  \\ \nonumber\\
\textrm{(WC-SBS)} \qquad  \qquad & \left[ \partial_{\tilde{t}} + \tilde{\nu}_3^B \right] n_3 = - \frac{1}{4}  \frac{Zm_e}{m_i} \frac{\omega_2}{\omega_3} \frac{c^2 k_3^2}{\omega_2^2} \left(a_1^* a_2\right) \label{eqn:SBS1} \\ \nonumber\\
\textrm{(SC-SBS)} \qquad  \qquad & \left[ \partial^2_{\tilde{t}} - 2i \frac{\omega_3}{\omega_2} \tilde{\nu}_3^B \right] n_3 = \frac{i}{2}  \frac{Zm_e}{m_i} \frac{c^2 k_3^2}{\omega_2^2} \left(a_1^* a_2\right)  \label{eqn:SBS2} \\ \nonumber\\
\textrm{(SBS)} \qquad  \qquad & \left[ \frac{i}{2} \frac{\omega_2}{\omega_3} \partial^2_{\tilde{t}}  + \partial_{\tilde{t}} + \tilde{\nu}_3^B \right] n_3 = -\frac{1}{4} \frac{Zm_e}{m_i} \frac{\omega_2}{\omega_3} \frac{c^2 k_3^2}{\omega_2^2} \left(a_1^* a_2\right)\label{eqn:SBS3} 
\end{align}
\end{widetext}
These equations can numerically, or, in some limits, analytically, be solved to describe amplification by SRS or SBS subject to the assumptions discussed above. 

For this model to be useful, particularly in the x-ray regime, the damping coefficients $\tilde{\nu}_{1,2,3}$, in the above equations need to be evaluated. We include collisional damping of light waves, Langmuir waves, and the ion response, as well as Landau damping of Langmuir waves. The collisional damping of electromagnetic waves is \cite{Ginzburg1970}:
\begin{equation}
\tilde{\nu}_{1,2} = \frac{\omega_{pe}^2}{\omega_{1,2}^2} \tilde{\nu}_{\textrm{ei}}
\end{equation}
where
\begin{equation}
\tilde{\nu}_{\textrm{ei}} = \frac{2 \sqrt{2}}{3 \sqrt{\pi}} \frac{\Lambda_C r_e \omega_{1,2}^3}{q_T^{3/2} c \omega_2 \omega_e}
\end{equation}
Here, $q_T = 4T_e/Nm_e c^2$, $\Lambda_C = \ln 12 \pi \lambda_D^3 n_e /Z$ is the Coulomb logarithm, $\lambda_D = (T_e/4\pi n_e e^2)^{1/2}$ is the Debye length and $r_e = e^2/m_e c^2$.

The Langmuir wave is both Landau damped and collisionally damped, so, $\tilde{\nu}_3 = \tilde{\nu}_{\textrm{Lnd}} + \tilde{\nu}_{\textrm{ei}}/4$ \cite{Malkin2009quasitransient}, where:
\begin{equation}
\tilde{\nu}_{\textrm{Lnd}} = \frac{\sqrt{N \pi}}{(2q_T)^{3/2}} e^{-\frac{1}{q_T}}
\label{eqn:landamp2}
\end{equation}
Collisional damping of the ion-acoustic wave has a smaller effect, which has been approximated as $\tilde{\nu}_3 = 0.01 \times 2c_s/c$ \cite{Epperlein1992Damping}. Since this damping rate is small compared to the direct collisional damping rate for light waves, its exact value does not substantially affect observed behavior. We neglect Landau damping of the ion-acoustic wave by staying in the regime $T_e \gg T_i$. 

\section{The Asymptotic Linear Growth Rate}
The growth rate of the seed in the linear regime, where the pump is not significantly depleted, is important for determining the conditions under which plasma amplification is possible. An exact solution for the impulse response of the seed-plasma wave system, which is applicable to SRS and SBS amplification for negligible pump damping, has been derived by Bobroff and Haus \cite{Bobroff1967impulse}, albeit with the wave equations cast in slightly different form than that usually applied for plasma amplification. Using this method, Malkin and Fisch \cite{Malkin2009quasitransient} found the effective linear growth rate of the seed in the quasi-transient backward Raman amplification (QBRA) regime, adjusted for damping of the plasma wave. This growth rate is the asymptotic solution at infinite time, and represents the rate at which the seed maximum will grow.

Here, we use the self-similarity of the asymptotic solution to find the effective growth of the seed pulse, a method which is simple and more readily extended to include direct damping of the seed and strongly-coupled SBS. Starting with the equations applicable to SRS and WC-SBS, we assume that the solution in the limit $t \to \infty$ of the coupled equations:
\begin{align}
\left[ \partial_{\tilde{t}} + \tilde{v}_1 \partial_{\tilde{x}} + \tilde{\nu}_1 \right] a_1 &= K_1 n_3 \\
\left[ \partial_{\tilde{t}} + \tilde{\nu}_3 \right] n_3 &= K_3 a_1
\end{align}
is of the form 
\begin{equation}
a_1(z,t) = \alpha (\xi) e^{\kappa \tilde{t}} \quad \quad n_3(z,t) = \eta (\xi) e^{\kappa \tilde{t}}
\end{equation}
where
\begin{equation}
\xi = \frac{\tilde{x}}{v \tilde{t}}
\end{equation}
Note that this method includes damping of the seed ($\tilde{\nu}_1$) and plasma response ($\tilde{\nu}_3$), but does not include any depletion or damping of the pump. The constant $v$ represents the unknown rate at which the seed is stretching in time. Substituting the above forms of the solution,
\begin{align}
\left[ \partial_{\tilde{t}} + \frac{\tilde{v}_1}{v \tilde{t}} \partial_{\xi} + \tilde{\nu}_1 \right] \alpha (\xi) e^{\kappa \tilde{t}} &= K_1\eta (\xi) e^{\kappa \tilde{t}} \\
\left[ \partial_{\tilde{t}} + \tilde{\nu}_3 \right] \eta (\xi) e^{\kappa \tilde{t}} &= K_3 \alpha (\xi) e^{\kappa \tilde{t}}
\end{align}
which, eliminating the density, may be rewritten as a single equation:
\begin{equation}
\left[ \partial_{\tilde{t}} + \tilde{\nu}_3 \right] \left[ \partial_{\tilde{t}} + \frac{\tilde{v}_1}{v \tilde{t}} \partial_{\xi} + \tilde{\nu}_1 \right] \alpha e^{\kappa \tilde{t}} = K_1 K_3 \alpha e^{\kappa \tilde{t}}
\end{equation}
Evaluating the derivatives, we have an equation with which we want to solve for $\kappa$, the long-time growth rate. 
\begin{align}
\kappa^2 \alpha + \tilde{\nu}_3 \kappa \alpha &- \frac{\tilde{v}_1}{v \tilde{t}^2} \partial_\xi \alpha + \frac{\tilde{v}_1}{v \tilde{t}} \kappa \partial_\xi \alpha  \alpha \nonumber \\*
& + \frac{\tilde{v}_1 \tilde{\nu}_3 }{v \tilde{t}} \partial_\xi + \kappa \tilde{\nu}_1 \alpha + \tilde{\nu}_3\tilde{\nu}_1 \alpha = K_1 K_3 \alpha
\end{align}
In the infinite time limit, we may drop terms with inverse powers of $\tilde{t}$, which removes all of the $\xi$-derivatives from the equation and leaves only a quadratic equation in $\kappa$.
\begin{equation}
\kappa^2 + \kappa \left( \tilde{\nu}_1 + \tilde{\nu}_3\right)  + \tilde{\nu}_3 \tilde{\nu}_1 - \Gamma_R^2 = 0
\end{equation}
This may be solved in straightforward fashion to give the modified growth rate:
\begin{equation}
\kappa = \frac{1}{2} \sqrt{4 K_1 K_3 + \left(\tilde{\nu}_3 - \tilde{\nu}_1 \right)^2} - \frac{\tilde{\nu}_3 + \tilde{\nu}_1}{2}
\label{eqn:modgrowth2}
\end{equation}
which for SRS is:
\begin{equation}
\hat{\Gamma}_R = \frac{1}{2} \sqrt{4 \tilde{\Gamma}_R^2 + \left(\tilde{\nu}_3 - \tilde{\nu}_1 \right)^2} - \frac{\tilde{\nu}_3 + \tilde{\nu}_1}{2}
\end{equation}
where $\tilde{\Gamma}_R = \Gamma_R/\omega_2$ is the undamped Raman growth rate. The above expression will also be valid for WC-SBS, substituting $\Gamma_{B_{wc}}$ for $\Gamma_R$ and using the appropriate damping factors. Although this form is different from that presented by Malkin and Fisch \cite{Malkin2009quasitransient}, if $\tilde{\nu}_1 = 0$, it is strictly equivalent.

To treat SBS more generally, including the strongly-coupled and intermediate cases, we return to the wave-coupling equations which include the second derivative in time associated with strongly-coupled stimulated Brillouin scattering:
\begin{align}
\left[ \partial_{\tilde{t}} + \tilde{v}_1 \partial_{\tilde{x}} + \tilde{\nu}_1 \right] a_1 &= K_1 n_3 \\
\left[  C_{sc} \partial^2_{\tilde{t}}  + C_{wc} \partial_{\tilde{t}} + \tilde{\nu}_3 \right] n_3 &= K_3 a_1
\end{align}
where for SBS
\begin{equation}
C_{sc} = \frac{i}{2} \frac{\omega_2}{\omega_3}
\end{equation}
and $C_{wc} = 1$. For WC-SBS, we can assume $C_{sc} = 0$ and for SC-SBS, we assume $C_{wc} = 0$. Similarly to the previous derivation, the derivatives with respect to $\xi$ vanish as $\tilde{t} \to \infty$, leaving the following cubic equation for $\kappa$:
\begin{align}
C_{sc} \kappa^3 +\left(C_{sc} \tilde{\nu}_1 + C_{wc}\right) \kappa^2  + \left(C_{wc} \tilde{\nu}_1  +\tilde{\nu}_3\right) \kappa & \nonumber \\
 +\: \tilde{\nu}_1 \tilde{\nu}_3 - K_1K_3 &= 0
\end{align}
Since this is a cubic equation in $\kappa$, a solution exists, though not one in as simple a form as that available for SRS. We give the full analytic solution below, though note that, in practice, numerical evaluation of the above equation is likely more readily implemented:
\begin{equation}
\kappa = \max \left( \textrm{Re} \left[ -\frac{1}{3A} \left(B + G + \frac{E}{G} \right) \right] \right)
\end{equation}
where
\begin{align}
A &= C_{sc} \nonumber \\
B & = C_{sc} \tilde{\nu}_1 + C_{wc} \nonumber\\
C & = C_{wc} \tilde{\nu}_1 + \tilde{\nu}_3 \nonumber\\
D &= \tilde{\nu}_1\tilde{\nu}_3 - K_1 K_3 \\
E & = B^2 - 3AC \nonumber\\
F & = 2B^3 - 9ABC +27A^2D \nonumber\\
G &= \sqrt[3]{\frac{F \pm \sqrt{F^2 - 4E^3}}{2}} \nonumber
\end{align}
When $C_{sc} = 0$, Eq.~(\ref{eqn:modgrowth2}) should be used directly. When $\tilde{\nu}_1$ and $\tilde{\nu}_3$ are zero, this equation recovers the plane wave growth rate \cite{Edwards2016short}. 

%
%
%

\section{Steady-State Solution for Non-Linear Amplification}
With the inclusion of collisional damping of the seed laser pulse ($\tilde{\nu}_1$), the solution to the three-wave equations in the pump-depletion (non-linear) regime asymptotically approaches a steady state at infinite time, rather than growing infinitely short and intense (until violation of the envelope approximation invalidates the SRS and SBS models). Although the steady-state solution is of limited utility in finding actual amplified pulse shapes because secondary effects, including relativistic self-modulation, group velocity dispersion, and forward Raman scattering, become increasingly important as the growth decreases, calculation of this asymptote allows us to set a limit on the maximum possible factor of amplification which can be achieved with a particular set of parameters. As we show, this limit is useful for identifying regimes for which plasma amplification is unlikely to be viable, but by itself, does not guarantee that a particular choice of parameters is practical. Since the time taken to reach the steady state solution as important as its final value, for some applications it may be desirable to use a varying plasma density, combining the high growth rates of higher density plasmas for early times with the high steady-state amplification factors of lower plasma densities for the final stages of interaction. With appropriate choice of parameters, the following analysis is valid for both SRS and all forms of SBS.

Starting with the three-wave equations written in the form:
\begin{align}
\left[ \partial_{\tilde{t}} + \tilde{v}_1 \partial_{\tilde{x}} + \tilde{\nu}_1 \right] a_1 &= K_1 \left(n_3^* a_2\right) \\ \nonumber\\
\left[ \partial_{\tilde{t}} + \tilde{v}_2 \partial_{\tilde{x}} + \tilde{\nu}_2 \right] a_2 &= K_2 \left(n_3 a_1\right) \\ \nonumber\\
\left[ \frac{i}{2} \frac{\omega_2}{\omega_3} \partial^2_{\tilde{t}}  + \partial_{\tilde{t}}  + \tilde{v}_3\partial_{\tilde{x}} + \tilde{\nu}_3 \right] n_3 &= K_3\left(a_1^* a_2\right) \label{eqn:ApC3}
\end{align}
where the first term on the left hand side of Eq.~(\ref{eqn:ApC3}) is immediately negligible for SRS and
\begin{align}
K_1 &= -\frac{1}{4}\frac{\omega_2}{\omega_1} N \\
K_2 &= \frac{1}{4} N \\
K_3^{R} &= - \frac{1}{4}  \frac{1}{\sqrt{N}} \frac{c^2 k_3^2}{\omega_2^2} \\
K_3^B &= -\frac{1}{4} \frac{Zm_e}{m_i} \frac{\omega_2}{\omega_3} \frac{c^2 k_3^2}{\omega_2^2}
\end{align}
we find the asymptotic solution by choosing the reference frame moving with the seed and by setting the time derivatives to zero. We will also neglect pump damping ($\tilde{\nu}_2$) because its inclusion causes the system to be propagation-length dependent (i.e. will not reach a steady state), and pump-depletion can be usefully included after this calculation using a pre-compensated variable-intensity pump boundary condition. These solutions therefore lose validity if the pump is substantially depleted on the timescale of the interaction, a limitation mitigated by the fact that a high degree of pump depletion will prevent pump-seed interaction in a plasma of any reasonable length, making such a regime unlikely to be useful. The steady-state equations, assuming $|\tilde{v}_1| \gg |\tilde{v}_3|$, are:
\begin{align}
\tilde{\nu}_1 a_1 &= K_1 n_3 a_2 \\
(\tilde{v}_2-\tilde{v}_1) \partial_{\tilde{x}} a_2 &= K_2 n_3 a_1 \\
-\tilde{v}_1  \partial_{\tilde{x}} n_3 + \tilde{\nu}_3 n_3 &= K_3 a_1 a_2 
\end{align}

In the case of SBS, damping of the ion-wave ($\tilde{\nu}_3$) is small and can be safely neglected, substantially simplifying the system of equations. Substituting for $a_1$, the above equations may be rewritten as:
\begin{align}
\partial_{\tilde{x}} a_2^2 &= 2 D_1 n_3^2 a_2^2 \\
\partial_{\tilde{x}} n_3^2 &= 2 D_2 n_3^2 a_2^2 
\end{align}
where
\begin{equation}
D_1 = \frac{K_2K_1}{(\tilde{v}_2 -\tilde{v}_1)\tilde{\nu}_1} \qquad D_2 =  \frac{-K_3 K_1}{\tilde{v}_1 \tilde{\nu}_1} 
\end{equation}
Note that $v_1 < 0$. This set of equations may be solved to give:
\begin{equation}
a_2(\tilde{x}) = a_{20} \left[ \frac{1}{1 + \frac{1}{2D_2} e^{2 D_2 a_{20}^2 \tilde{x}}} \right]^{\frac{1}{2}}
\end{equation}
We are, however, primarily interested in $a_1$, which may be written in terms of $a_2$ as:
\begin{equation}
a_1 = \frac{K_1}{\tilde{\nu}_1} a_2 n_3 = \frac{K_1}{\tilde{\nu}_3} \left[ \frac{1}{2D_1} \partial_{\tilde{x}} a_2^2 \right]^{\frac{1}{2}}
\end{equation}
This simplifies to:
\begin{equation}
a_1(\tilde{x}) = \frac{-K_1a_{20}^2}{\tilde{\nu}_1 \sqrt{2D_1}} \frac{e^{D_2 a_{20}^2 \tilde{x}}}{1+\frac{1}{2D_2}e^{2D_2 a_{20}^2 \tilde{x}}}
\end{equation}
To get the maximum value of $a_1$ in the steady state, we find where the derivative with respect to $\tilde{x}$ is zero to get:
\begin{equation}
a_1^{\textrm{max}} = \frac{-K_1 a_{2,0}^2}{2 \tilde{\nu}_1} \sqrt{\frac{-K_3}{K_2} \left(1-\frac{\tilde{v}_2}{\tilde{v}_1}\right)}
\end{equation}
Note that since $K_1$, $K_3$ and $\tilde{v}_1$ are less than 0, the above envelope is real and greater than 0. It is clear that if $\tilde{\nu}_1 = 0$, the maximum intensity determined from this set of equations is infinite, and they cannot be used to determine maximum possible amplification. If $\tilde{\nu}_3$ is not small, for example for SRS, the system of equations is less tractable, and we resort to numerical evaluation of the steady state equations. The results of evaluating these equations are shown in Fig.~4.

\section{Numerical Solutions to the Coupling Equations}
To numerically solve the wave-coupling equations, we employ a step-shifting algorithm similar to that previously applied to the problem of Raman amplification by Clark \cite{ClarkThesis,Clark2003operating,Clark2003particle}. To implement the propagators on the left side of Eqs.~(1)-(4), all simulation values are shifted by one grid space at intervals of $\Delta x/\tilde{v}_{(1,2,3)}$. The time derivatives are evaluated with a fourth-order Runge-Kutta scheme. Numerical evaluation of the coupling equations is much faster than fully kinetic or fluid based simulations, and provides solutions for conditions which are not analytically tractable. As code validation, and to justify comments made in the manuscript, we present a series of simple comparisons between simulations and analytic results. 

\begin{figure}
\centering
\includegraphics[width=\linewidth]{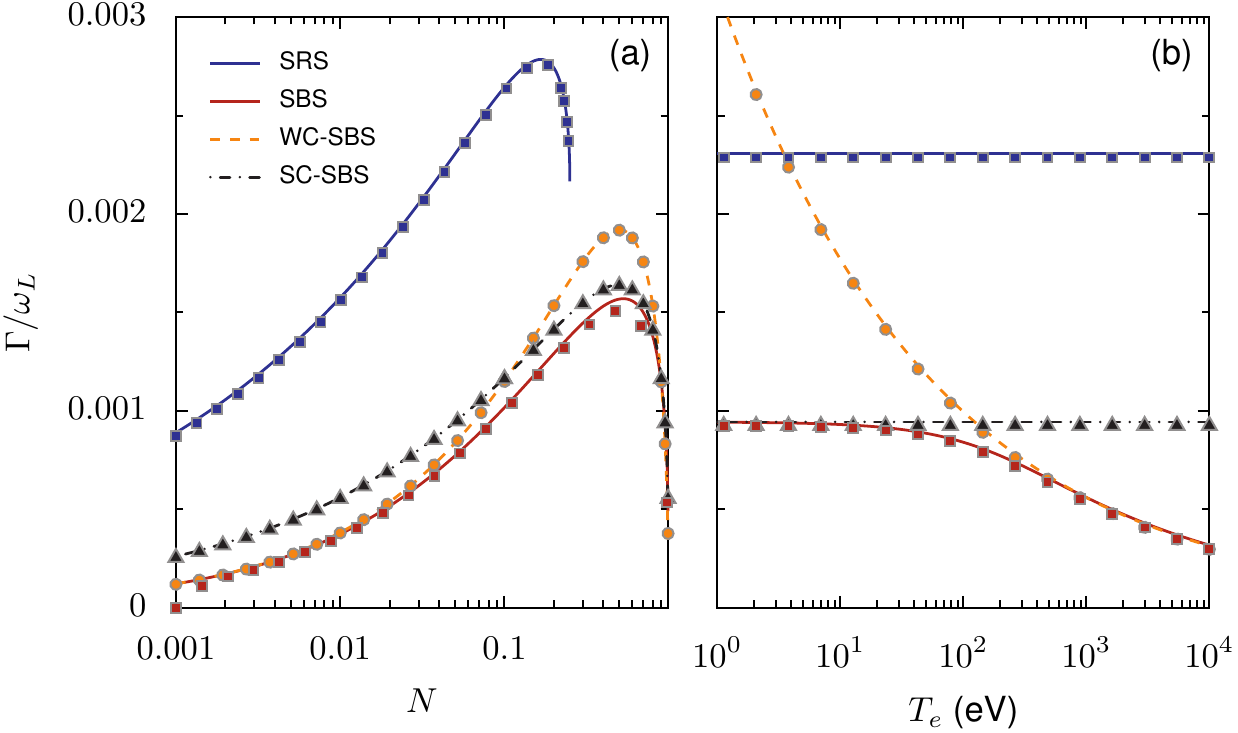}
\caption{Undamped linear growth rate of SRS and SBS as a function of (a) density (with $T_e = 200$ eV) and (b) temperature (at $N = 0.05$) at fixed pump strength ($a_0 = 0.01$) and wavelength ($\lambda = 1$ $\mu$m). The analytic linear growth rates (lines) are compared to those extracted from the rate of growth of the maximum of the seed in wave-coupling calculations.}
\label{fig:AD1}
\end{figure}

The linear growth rate of the undamped system represents the early-time behavior of the interaction and is important for quantifying the highest degree of amplification that may be expected. It can be shown that the undamped asymptotic seed growth rate in the linear regime is simply the instability plane-wave growth rate, for which expressions are well known for SRS, WC-SBS, and SC-SBS and which may be readily calculated for intermediate SBS conditions \cite{Edwards2016short}. In Fig.~\ref{fig:AD1}, the analytic growth rate  for the undamped SRS and SBS instabilities are compared to the asymptotic linear-regime growth rates found from numerical solution of the three wave-coupling equations. Note that, for example, a WC-SBS solution may be found in the strongly-coupled regime, but this solution will disagree with the solution to the full equation and does not represent a useful solution to the full system. Furthermore, as evaluated here, the equations assume that the regime of interaction is set only by the initial boundary field strength, and do not fully include the transition between regimes that would be observed in practice \cite{Schluck2016dynamical}. 

\begin{figure}
\centering
\includegraphics[width=\linewidth]{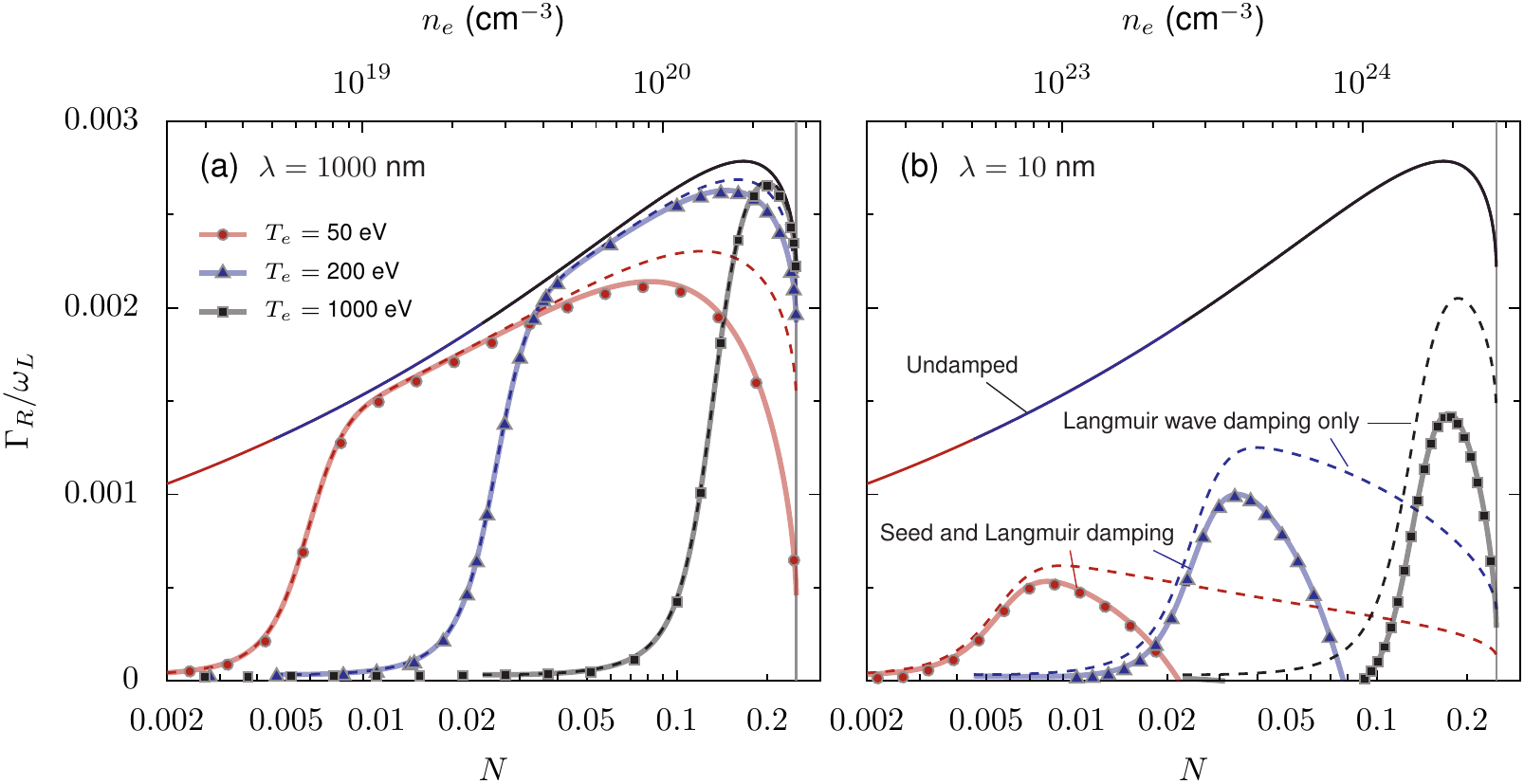}
\caption{Analytic estimates of the linear growth rate of a Raman amplified seed in the presence of no damping (solid thin lines), Landau and collisional damping of only the plasma wave (dashed lines), and both damping of the plasma wave and collisional damping of the seed (thick solid lines), compared to the growth rate extracted three-wave simulations (points) with both seed and plasma wave damping. Simulations are shown for $T_e = 50$, $200$, and $1000$ eV, and at (a) $\lambda = 1000$ nm and (b) $10$ nm. All results are for $a_0 = 0.01$.}
\label{fig:AD2}
\end{figure}

The expressions for the damped growth rate in the linear regime [Eqs.~(6) and (7)] may also be compared to the reduced growth rate found using numerical evaluation of the three-wave model. In Fig.~\ref{fig:AD1}, the analytic solutions are shown for SRS with no damping, damping of only the Langmuir wave, and damping of both the Langmuir and seed electromagnetic wave. These are compared to three-wave calculations which include both Langmuir wave and seed damping. As should be expected, the numerical solutions agree with the analytic model. Similarly, Fig.~\ref{fig:AD1}, compares analytic solutions and numerical solutions for the damped growth rate of SBS, using the weakly-coupled, strongly-coupled, and full models. The agreement suggests the validity of the analytic expressions and the numerical implementation.


\begin{figure}[t]
\centering
\includegraphics[width=\linewidth]{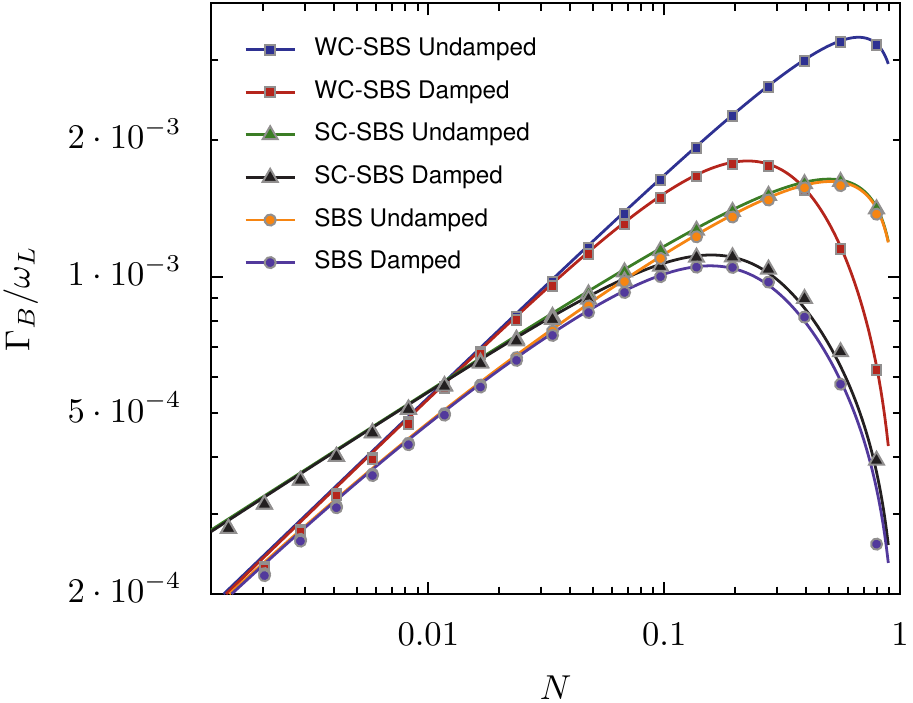}
\caption{Analytic estimates (lines) compared to values found from solving the three-wave equations (points) for the Brillouin linear growth rate in the WC-SBS, SC-SBS, and SBS models with and without damping (which arises almost entirely due to collisional damping of the seed) at $\lambda = 1000$ nm, $a_0 = 0.01$, and $T_e = 50$ eV. The ion temperature is assumed negligible, $m_i = 1836 m_e$, and $Z = 1$.}
\label{fig:AD3}
\end{figure}


\begin{thebibliography}{82}%
\makeatletter
\providecommand \@ifxundefined [1]{%
 \@ifx{#1\undefined}
}%
\providecommand \@ifnum [1]{%
 \ifnum #1\expandafter \@firstoftwo
 \else \expandafter \@secondoftwo
 \fi
}%
\providecommand \@ifx [1]{%
 \ifx #1\expandafter \@firstoftwo
 \else \expandafter \@secondoftwo
 \fi
}%
\providecommand \natexlab [1]{#1}%
\providecommand \enquote  [1]{``#1''}%
\providecommand \bibnamefont  [1]{#1}%
\providecommand \bibfnamefont [1]{#1}%
\providecommand \citenamefont [1]{#1}%
\providecommand \href@noop [0]{\@secondoftwo}%
\providecommand \href [0]{\begingroup \@sanitize@url \@href}%
\providecommand \@href[1]{\@@startlink{#1}\@@href}%
\providecommand \@@href[1]{\endgroup#1\@@endlink}%
\providecommand \@sanitize@url [0]{\catcode `\\12\catcode `\$12\catcode
  `\&12\catcode `\#12\catcode `\^12\catcode `\_12\catcode `\%12\relax}%
\providecommand \@@startlink[1]{}%
\providecommand \@@endlink[0]{}%
\providecommand \url  [0]{\begingroup\@sanitize@url \@url }%
\providecommand \@url [1]{\endgroup\@href {#1}{\urlprefix }}%
\providecommand \urlprefix  [0]{URL }%
\providecommand \Eprint [0]{\href }%
\providecommand \doibase [0]{http://dx.doi.org/}%
\providecommand \selectlanguage [0]{\@gobble}%
\providecommand \bibinfo  [0]{\@secondoftwo}%
\providecommand \bibfield  [0]{\@secondoftwo}%
\providecommand \translation [1]{[#1]}%
\providecommand \BibitemOpen [0]{}%
\providecommand \bibitemStop [0]{}%
\providecommand \bibitemNoStop [0]{.\EOS\space}%
\providecommand \EOS [0]{\spacefactor3000\relax}%
\providecommand \BibitemShut  [1]{\csname bibitem#1\endcsname}%
\let\auto@bib@innerbib\@empty
\bibitem [{\citenamefont {Ackermann}\ \emph {et~al.}(2007)\citenamefont
  {Ackermann}, \citenamefont {Asova}, \citenamefont {Ayvazyan}, \citenamefont
  {Azima}, \citenamefont {Baboi}, \citenamefont {B{\"a}hr}, \citenamefont
  {Balandin}, \citenamefont {Beutner}, \citenamefont {Brandt}, \citenamefont
  {Bolzmann} \emph {et~al.}}]{Ackermann2007operation}%
  \BibitemOpen
  \bibfield  {author} {\bibinfo {author} {\bibfnamefont {W.}~\bibnamefont
  {Ackermann}}, \bibinfo {author} {\bibfnamefont {G.}~\bibnamefont {Asova}},
  \bibinfo {author} {\bibfnamefont {V.}~\bibnamefont {Ayvazyan}}, \bibinfo
  {author} {\bibfnamefont {A.}~\bibnamefont {Azima}}, \bibinfo {author}
  {\bibfnamefont {N.}~\bibnamefont {Baboi}}, \bibinfo {author} {\bibfnamefont
  {J.}~\bibnamefont {B{\"a}hr}}, \bibinfo {author} {\bibfnamefont
  {V.}~\bibnamefont {Balandin}}, \bibinfo {author} {\bibfnamefont
  {B.}~\bibnamefont {Beutner}}, \bibinfo {author} {\bibfnamefont
  {A.}~\bibnamefont {Brandt}}, \bibinfo {author} {\bibfnamefont
  {A.}~\bibnamefont {Bolzmann}},  \emph {et~al.},\ }\bibfield  {title}
  {\bibinfo {title} {Operation of a free-electron laser from the extreme
  ultraviolet to the water window},\ }\href@noop {} {\bibfield  {journal}
  {\bibinfo  {journal} {Nature Photon.}\ }\textbf {\bibinfo {volume} {1}},\
  \bibinfo {pages} {336} (\bibinfo {year} {2007})}\BibitemShut {NoStop}%
\bibitem [{\citenamefont {Emma}\ \emph {et~al.}(2010)\citenamefont {Emma},
  \citenamefont {Akre}, \citenamefont {Arthur}, \citenamefont {Bionta},
  \citenamefont {Bostedt}, \citenamefont {Bozek}, \citenamefont {Brachmann},
  \citenamefont {Bucksbaum}, \citenamefont {Coffee}, \citenamefont {Decker}
  \emph {et~al.}}]{Emma2010first}%
  \BibitemOpen
  \bibfield  {author} {\bibinfo {author} {\bibfnamefont {P.}~\bibnamefont
  {Emma}}, \bibinfo {author} {\bibfnamefont {R.}~\bibnamefont {Akre}}, \bibinfo
  {author} {\bibfnamefont {J.}~\bibnamefont {Arthur}}, \bibinfo {author}
  {\bibfnamefont {R.}~\bibnamefont {Bionta}}, \bibinfo {author} {\bibfnamefont
  {C.}~\bibnamefont {Bostedt}}, \bibinfo {author} {\bibfnamefont
  {J.}~\bibnamefont {Bozek}}, \bibinfo {author} {\bibfnamefont
  {A.}~\bibnamefont {Brachmann}}, \bibinfo {author} {\bibfnamefont
  {P.}~\bibnamefont {Bucksbaum}}, \bibinfo {author} {\bibfnamefont
  {R.}~\bibnamefont {Coffee}}, \bibinfo {author} {\bibfnamefont {F.-J.}\
  \bibnamefont {Decker}},  \emph {et~al.},\ }\bibfield  {title} {\bibinfo
  {title} {First lasing and operation of an {\aa}ngstrom-wavelength
  free-electron laser},\ }\href@noop {} {\bibfield  {journal} {\bibinfo
  {journal} {Nature Photon.}\ }\textbf {\bibinfo {volume} {4}},\ \bibinfo
  {pages} {641} (\bibinfo {year} {2010})}\BibitemShut {NoStop}%
\bibitem [{\citenamefont {Fuchs}\ \emph {et~al.}(2015)\citenamefont {Fuchs},
  \citenamefont {Trigo}, \citenamefont {Chen}, \citenamefont {Ghimire},
  \citenamefont {Shwartz}, \citenamefont {Kozina}, \citenamefont {Jiang},
  \citenamefont {Henighan}, \citenamefont {Bray}, \citenamefont {Ndabashimiye}
  \emph {et~al.}}]{Fuchs2015anomalous}%
  \BibitemOpen
  \bibfield  {author} {\bibinfo {author} {\bibfnamefont {M.}~\bibnamefont
  {Fuchs}}, \bibinfo {author} {\bibfnamefont {M.}~\bibnamefont {Trigo}},
  \bibinfo {author} {\bibfnamefont {J.}~\bibnamefont {Chen}}, \bibinfo {author}
  {\bibfnamefont {S.}~\bibnamefont {Ghimire}}, \bibinfo {author} {\bibfnamefont
  {S.}~\bibnamefont {Shwartz}}, \bibinfo {author} {\bibfnamefont
  {M.}~\bibnamefont {Kozina}}, \bibinfo {author} {\bibfnamefont
  {M.}~\bibnamefont {Jiang}}, \bibinfo {author} {\bibfnamefont
  {T.}~\bibnamefont {Henighan}}, \bibinfo {author} {\bibfnamefont
  {C.}~\bibnamefont {Bray}}, \bibinfo {author} {\bibfnamefont {G.}~\bibnamefont
  {Ndabashimiye}},  \emph {et~al.},\ }\bibfield  {title} {\bibinfo {title}
  {{Anomalous nonlinear X-ray Compton scattering}},\ }\href@noop {} {\bibfield
  {journal} {\bibinfo  {journal} {Nature Phys.}\ }\textbf {\bibinfo {volume}
  {11}},\ \bibinfo {pages} {964} (\bibinfo {year} {2015})}\BibitemShut
  {NoStop}%
\bibitem [{\citenamefont {Krausz}\ and\ \citenamefont
  {Ivanov}(2009)}]{Krausz2009attosecond}%
  \BibitemOpen
  \bibfield  {author} {\bibinfo {author} {\bibfnamefont {F.}~\bibnamefont
  {Krausz}}\ and\ \bibinfo {author} {\bibfnamefont {M.}~\bibnamefont
  {Ivanov}},\ }\bibfield  {title} {\bibinfo {title} {Attosecond physics},\
  }\href@noop {} {\bibfield  {journal} {\bibinfo  {journal} {Rev. Mod. Phys.}\
  }\textbf {\bibinfo {volume} {81}},\ \bibinfo {pages} {163} (\bibinfo {year}
  {2009})}\BibitemShut {NoStop}%
\bibitem [{\citenamefont {Neutze}\ \emph {et~al.}(2000)\citenamefont {Neutze},
  \citenamefont {Wouts}, \citenamefont {van~der Spoel}, \citenamefont
  {Weckert},\ and\ \citenamefont {Hajdu}}]{Neutze2000potential}%
  \BibitemOpen
  \bibfield  {author} {\bibinfo {author} {\bibfnamefont {R.}~\bibnamefont
  {Neutze}}, \bibinfo {author} {\bibfnamefont {R.}~\bibnamefont {Wouts}},
  \bibinfo {author} {\bibfnamefont {D.}~\bibnamefont {van~der Spoel}}, \bibinfo
  {author} {\bibfnamefont {E.}~\bibnamefont {Weckert}}, \ and\ \bibinfo
  {author} {\bibfnamefont {J.}~\bibnamefont {Hajdu}},\ }\bibfield  {title}
  {\bibinfo {title} {Potential for biomolecular imaging with femtosecond x-ray
  pulses},\ }\href@noop {} {\bibfield  {journal} {\bibinfo  {journal} {Nature}\
  }\textbf {\bibinfo {volume} {406}},\ \bibinfo {pages} {752} (\bibinfo {year}
  {2000})}\BibitemShut {NoStop}%
\bibitem [{\citenamefont {Miao}\ \emph {et~al.}(2015)\citenamefont {Miao},
  \citenamefont {Ishikawa}, \citenamefont {Robinson},\ and\ \citenamefont
  {Murnane}}]{Miao2015beyond}%
  \BibitemOpen
  \bibfield  {author} {\bibinfo {author} {\bibfnamefont {J.}~\bibnamefont
  {Miao}}, \bibinfo {author} {\bibfnamefont {T.}~\bibnamefont {Ishikawa}},
  \bibinfo {author} {\bibfnamefont {I.~K.}\ \bibnamefont {Robinson}}, \ and\
  \bibinfo {author} {\bibfnamefont {M.~M.}\ \bibnamefont {Murnane}},\
  }\bibfield  {title} {\bibinfo {title} {{Beyond crystallography: Diffractive
  imaging using coherent x-ray light sources}},\ }\href@noop {} {\bibfield
  {journal} {\bibinfo  {journal} {Science}\ }\textbf {\bibinfo {volume}
  {348}},\ \bibinfo {pages} {530} (\bibinfo {year} {2015})}\BibitemShut
  {NoStop}%
\bibitem [{\citenamefont {Bostedt}\ \emph {et~al.}(2016)\citenamefont
  {Bostedt}, \citenamefont {Boutet}, \citenamefont {Fritz}, \citenamefont
  {Huang}, \citenamefont {Lee}, \citenamefont {Lemke}, \citenamefont {Robert},
  \citenamefont {Schlotter}, \citenamefont {Turner},\ and\ \citenamefont
  {Williams}}]{Bostedt2016linac}%
  \BibitemOpen
  \bibfield  {author} {\bibinfo {author} {\bibfnamefont {C.}~\bibnamefont
  {Bostedt}}, \bibinfo {author} {\bibfnamefont {S.}~\bibnamefont {Boutet}},
  \bibinfo {author} {\bibfnamefont {D.~M.}\ \bibnamefont {Fritz}}, \bibinfo
  {author} {\bibfnamefont {Z.}~\bibnamefont {Huang}}, \bibinfo {author}
  {\bibfnamefont {H.~J.}\ \bibnamefont {Lee}}, \bibinfo {author} {\bibfnamefont
  {H.~T.}\ \bibnamefont {Lemke}}, \bibinfo {author} {\bibfnamefont
  {A.}~\bibnamefont {Robert}}, \bibinfo {author} {\bibfnamefont {W.~F.}\
  \bibnamefont {Schlotter}}, \bibinfo {author} {\bibfnamefont {J.~J.}\
  \bibnamefont {Turner}}, \ and\ \bibinfo {author} {\bibfnamefont {G.~J.}\
  \bibnamefont {Williams}},\ }\bibfield  {title} {\bibinfo {title} {{Linac
  Coherent Light Source: The first five years}},\ }\href@noop {} {\bibfield
  {journal} {\bibinfo  {journal} {Rev. Mod. Phys.}\ }\textbf {\bibinfo {volume}
  {88}},\ \bibinfo {pages} {015007} (\bibinfo {year} {2016})}\BibitemShut
  {NoStop}%
\bibitem [{\citenamefont {Schwinger}(1951)}]{Schwinger1951gauge}%
  \BibitemOpen
  \bibfield  {author} {\bibinfo {author} {\bibfnamefont {J.}~\bibnamefont
  {Schwinger}},\ }\bibfield  {title} {\bibinfo {title} {On gauge invariance and
  vacuum polarization},\ }\href@noop {} {\bibfield  {journal} {\bibinfo
  {journal} {Phys. Rev.}\ }\textbf {\bibinfo {volume} {82}},\ \bibinfo {pages}
  {664} (\bibinfo {year} {1951})}\BibitemShut {NoStop}%
\bibitem [{\citenamefont {Brezin}\ and\ \citenamefont
  {Itzykson}(1970)}]{Brezin1970pair}%
  \BibitemOpen
  \bibfield  {author} {\bibinfo {author} {\bibfnamefont {E.}~\bibnamefont
  {Brezin}}\ and\ \bibinfo {author} {\bibfnamefont {C.}~\bibnamefont
  {Itzykson}},\ }\bibfield  {title} {\bibinfo {title} {Pair production in
  vacuum by an alternating field},\ }\href@noop {} {\bibfield  {journal}
  {\bibinfo  {journal} {Phys. Rev. D}\ }\textbf {\bibinfo {volume} {2}},\
  \bibinfo {pages} {1191} (\bibinfo {year} {1970})}\BibitemShut {NoStop}%
\bibitem [{\citenamefont {Popov}(1971)}]{Popov1971production}%
  \BibitemOpen
  \bibfield  {author} {\bibinfo {author} {\bibfnamefont {V.~S.}\ \bibnamefont
  {Popov}},\ }\bibfield  {title} {\bibinfo {title} {Production of {e$^+$e$^-$}
  pairs in an alternating external field},\ }\href@noop {} {\bibfield
  {journal} {\bibinfo  {journal} {JETP Lett.}\ }\textbf {\bibinfo {volume}
  {13}},\ \bibinfo {pages} {185} (\bibinfo {year} {1971})}\BibitemShut
  {NoStop}%
\bibitem [{\citenamefont {Ringwald}(2001)}]{Ringwald2001pair}%
  \BibitemOpen
  \bibfield  {author} {\bibinfo {author} {\bibfnamefont {A.}~\bibnamefont
  {Ringwald}},\ }\bibfield  {title} {\bibinfo {title} {Pair production from
  vacuum at the focus of an x-ray free electron laser},\ }\href@noop {}
  {\bibfield  {journal} {\bibinfo  {journal} {Phys. Lett. B}\ }\textbf
  {\bibinfo {volume} {510}},\ \bibinfo {pages} {107} (\bibinfo {year}
  {2001})}\BibitemShut {NoStop}%
\bibitem [{\citenamefont {Alkofer}\ \emph {et~al.}(2001)\citenamefont
  {Alkofer}, \citenamefont {Hecht}, \citenamefont {Roberts}, \citenamefont
  {Schmidt},\ and\ \citenamefont {Vinnik}}]{Alkofer2001pair}%
  \BibitemOpen
  \bibfield  {author} {\bibinfo {author} {\bibfnamefont {R.}~\bibnamefont
  {Alkofer}}, \bibinfo {author} {\bibfnamefont {M.~B.}\ \bibnamefont {Hecht}},
  \bibinfo {author} {\bibfnamefont {C.~D.}\ \bibnamefont {Roberts}}, \bibinfo
  {author} {\bibfnamefont {S.~M.}\ \bibnamefont {Schmidt}}, \ and\ \bibinfo
  {author} {\bibfnamefont {D.~V.}\ \bibnamefont {Vinnik}},\ }\bibfield  {title}
  {\bibinfo {title} {Pair creation and an x-ray free electron laser},\
  }\href@noop {} {\bibfield  {journal} {\bibinfo  {journal} {Phys. Rev. Lett.}\
  }\textbf {\bibinfo {volume} {87}},\ \bibinfo {pages} {193902} (\bibinfo
  {year} {2001})}\BibitemShut {NoStop}%
\bibitem [{\citenamefont {Popov}(2001)}]{Popov2001schwinger}%
  \BibitemOpen
  \bibfield  {author} {\bibinfo {author} {\bibfnamefont {V.~S.}\ \bibnamefont
  {Popov}},\ }\bibfield  {title} {\bibinfo {title} {Schwinger mechanism of
  electron-positron pair production by the field of optical and {X}-ray lasers
  in vacuum},\ }\href@noop {} {\bibfield  {journal} {\bibinfo  {journal} {JETP
  Lett.}\ }\textbf {\bibinfo {volume} {74}},\ \bibinfo {pages} {133} (\bibinfo
  {year} {2001})}\BibitemShut {NoStop}%
\bibitem [{\citenamefont {Malkin}\ \emph {et~al.}(2007)\citenamefont {Malkin},
  \citenamefont {Fisch},\ and\ \citenamefont
  {Wurtele}}]{Malkin2007compression}%
  \BibitemOpen
  \bibfield  {author} {\bibinfo {author} {\bibfnamefont {V.~M.}\ \bibnamefont
  {Malkin}}, \bibinfo {author} {\bibfnamefont {N.~J.}\ \bibnamefont {Fisch}}, \
  and\ \bibinfo {author} {\bibfnamefont {J.~S.}\ \bibnamefont {Wurtele}},\
  }\bibfield  {title} {\bibinfo {title} {{Compression of powerful x-ray pulses
  to attosecond durations by stimulated Raman backscattering in plasmas}},\
  }\href@noop {} {\bibfield  {journal} {\bibinfo  {journal} {Phys. Rev. E}\
  }\textbf {\bibinfo {volume} {75}},\ \bibinfo {pages} {026404} (\bibinfo
  {year} {2007})}\BibitemShut {NoStop}%
\bibitem [{\citenamefont {Sebban}\ \emph {et~al.}(2000)\citenamefont {Sebban},
  \citenamefont {Daido}, \citenamefont {Sakaya}, \citenamefont {Kato},
  \citenamefont {Murai}, \citenamefont {Tang}, \citenamefont {Gu},
  \citenamefont {Huang}, \citenamefont {Wang}, \citenamefont {Klisnick} \emph
  {et~al.}}]{Sebban2000full}%
  \BibitemOpen
  \bibfield  {author} {\bibinfo {author} {\bibfnamefont {S.}~\bibnamefont
  {Sebban}}, \bibinfo {author} {\bibfnamefont {H.}~\bibnamefont {Daido}},
  \bibinfo {author} {\bibfnamefont {N.}~\bibnamefont {Sakaya}}, \bibinfo
  {author} {\bibfnamefont {Y.}~\bibnamefont {Kato}}, \bibinfo {author}
  {\bibfnamefont {K.}~\bibnamefont {Murai}}, \bibinfo {author} {\bibfnamefont
  {H.}~\bibnamefont {Tang}}, \bibinfo {author} {\bibfnamefont {Y.}~\bibnamefont
  {Gu}}, \bibinfo {author} {\bibfnamefont {G.}~\bibnamefont {Huang}}, \bibinfo
  {author} {\bibfnamefont {S.}~\bibnamefont {Wang}}, \bibinfo {author}
  {\bibfnamefont {A.}~\bibnamefont {Klisnick}},  \emph {et~al.},\ }\bibfield
  {title} {\bibinfo {title} {Full characterization of a high-gain saturated
  x-ray laser at 13.9 nm},\ }\href@noop {} {\bibfield  {journal} {\bibinfo
  {journal} {Phys. Rev. A}\ }\textbf {\bibinfo {volume} {61}},\ \bibinfo
  {pages} {043810} (\bibinfo {year} {2000})}\BibitemShut {NoStop}%
\bibitem [{\citenamefont {Zeitoun}\ \emph {et~al.}(2004)\citenamefont
  {Zeitoun}, \citenamefont {Faivre}, \citenamefont {Sebban}, \citenamefont
  {Mocek}, \citenamefont {Hallou}, \citenamefont {Fajardo}, \citenamefont
  {Aubert}, \citenamefont {Balcou}, \citenamefont {Burgy}, \citenamefont
  {Douillet} \emph {et~al.}}]{Zeitoun2004high}%
  \BibitemOpen
  \bibfield  {author} {\bibinfo {author} {\bibfnamefont {{\relax
  Ph}.}~\bibnamefont {Zeitoun}}, \bibinfo {author} {\bibfnamefont
  {G.}~\bibnamefont {Faivre}}, \bibinfo {author} {\bibfnamefont
  {S.}~\bibnamefont {Sebban}}, \bibinfo {author} {\bibfnamefont
  {T.}~\bibnamefont {Mocek}}, \bibinfo {author} {\bibfnamefont
  {A.}~\bibnamefont {Hallou}}, \bibinfo {author} {\bibfnamefont
  {M.}~\bibnamefont {Fajardo}}, \bibinfo {author} {\bibfnamefont
  {D.}~\bibnamefont {Aubert}}, \bibinfo {author} {\bibfnamefont
  {P.}~\bibnamefont {Balcou}}, \bibinfo {author} {\bibfnamefont
  {F.}~\bibnamefont {Burgy}}, \bibinfo {author} {\bibfnamefont
  {D.}~\bibnamefont {Douillet}},  \emph {et~al.},\ }\bibfield  {title}
  {\bibinfo {title} {A high-intensity highly coherent soft {X-ray} femtosecond
  laser seeded by a high harmonic beam},\ }\href@noop {} {\bibfield  {journal}
  {\bibinfo  {journal} {Nature}\ }\textbf {\bibinfo {volume} {431}},\ \bibinfo
  {pages} {426} (\bibinfo {year} {2004})}\BibitemShut {NoStop}%
\bibitem [{\citenamefont {Suckewer}\ and\ \citenamefont
  {Jaegle}(2009)}]{Suckewer2009xray}%
  \BibitemOpen
  \bibfield  {author} {\bibinfo {author} {\bibfnamefont {S.}~\bibnamefont
  {Suckewer}}\ and\ \bibinfo {author} {\bibfnamefont {P.}~\bibnamefont
  {Jaegle}},\ }\bibfield  {title} {\bibinfo {title} {X-ray laser: past,
  present, and future},\ }\href@noop {} {\bibfield  {journal} {\bibinfo
  {journal} {Las. Phys. Lett.}\ }\textbf {\bibinfo {volume} {6}},\ \bibinfo
  {pages} {411} (\bibinfo {year} {2009})}\BibitemShut {NoStop}%
\bibitem [{\citenamefont {Alessi}\ \emph {et~al.}(2011)\citenamefont {Alessi},
  \citenamefont {Wang}, \citenamefont {Luther}, \citenamefont {Yin},
  \citenamefont {Martz}, \citenamefont {Woolston}, \citenamefont {Liu},
  \citenamefont {Berrill},\ and\ \citenamefont {Rocca}}]{Alessi2011efficient}%
  \BibitemOpen
  \bibfield  {author} {\bibinfo {author} {\bibfnamefont {D.}~\bibnamefont
  {Alessi}}, \bibinfo {author} {\bibfnamefont {Y.}~\bibnamefont {Wang}},
  \bibinfo {author} {\bibfnamefont {B.~M.}\ \bibnamefont {Luther}}, \bibinfo
  {author} {\bibfnamefont {L.}~\bibnamefont {Yin}}, \bibinfo {author}
  {\bibfnamefont {D.~H.}\ \bibnamefont {Martz}}, \bibinfo {author}
  {\bibfnamefont {M.~R.}\ \bibnamefont {Woolston}}, \bibinfo {author}
  {\bibfnamefont {Y.}~\bibnamefont {Liu}}, \bibinfo {author} {\bibfnamefont
  {M.}~\bibnamefont {Berrill}}, \ and\ \bibinfo {author} {\bibfnamefont
  {J.~J.}\ \bibnamefont {Rocca}},\ }\bibfield  {title} {\bibinfo {title}
  {Efficient excitation of gain-saturated sub-9-nm-wavelength tabletop
  soft-{X}-ray lasers and lasing down to 7.36 nm},\ }\href@noop {} {\bibfield
  {journal} {\bibinfo  {journal} {Phys. Rev. X}\ }\textbf {\bibinfo {volume}
  {1}},\ \bibinfo {pages} {021023} (\bibinfo {year} {2011})}\BibitemShut
  {NoStop}%
\bibitem [{\citenamefont {Rohringer}\ \emph {et~al.}(2012)\citenamefont
  {Rohringer}, \citenamefont {Ryan}, \citenamefont {London}, \citenamefont
  {Purvis}, \citenamefont {Albert}, \citenamefont {Dunn}, \citenamefont
  {Bozek}, \citenamefont {Bostedt}, \citenamefont {Graf}, \citenamefont {Hill}
  \emph {et~al.}}]{Rohringer2012atomic}%
  \BibitemOpen
  \bibfield  {author} {\bibinfo {author} {\bibfnamefont {N.}~\bibnamefont
  {Rohringer}}, \bibinfo {author} {\bibfnamefont {D.}~\bibnamefont {Ryan}},
  \bibinfo {author} {\bibfnamefont {R.~A.}\ \bibnamefont {London}}, \bibinfo
  {author} {\bibfnamefont {M.}~\bibnamefont {Purvis}}, \bibinfo {author}
  {\bibfnamefont {F.}~\bibnamefont {Albert}}, \bibinfo {author} {\bibfnamefont
  {J.}~\bibnamefont {Dunn}}, \bibinfo {author} {\bibfnamefont {J.~D.}\
  \bibnamefont {Bozek}}, \bibinfo {author} {\bibfnamefont {C.}~\bibnamefont
  {Bostedt}}, \bibinfo {author} {\bibfnamefont {A.}~\bibnamefont {Graf}},
  \bibinfo {author} {\bibfnamefont {R.}~\bibnamefont {Hill}},  \emph {et~al.},\
  }\bibfield  {title} {\bibinfo {title} {Atomic inner-shell {X}-ray laser at
  1.46 nanometres pumped by an {X}-ray free-electron laser},\ }\href@noop {}
  {\bibfield  {journal} {\bibinfo  {journal} {Nature}\ }\textbf {\bibinfo
  {volume} {481}},\ \bibinfo {pages} {488} (\bibinfo {year}
  {2012})}\BibitemShut {NoStop}%
\bibitem [{\citenamefont {Burnett}\ \emph {et~al.}(1977)\citenamefont
  {Burnett}, \citenamefont {Baldis}, \citenamefont {Richardson},\ and\
  \citenamefont {Enright}}]{Burnett1977harmonic}%
  \BibitemOpen
  \bibfield  {author} {\bibinfo {author} {\bibfnamefont {N.~H.}\ \bibnamefont
  {Burnett}}, \bibinfo {author} {\bibfnamefont {H.~A.}\ \bibnamefont {Baldis}},
  \bibinfo {author} {\bibfnamefont {M.~C.}\ \bibnamefont {Richardson}}, \ and\
  \bibinfo {author} {\bibfnamefont {G.~D.}\ \bibnamefont {Enright}},\
  }\bibfield  {title} {\bibinfo {title} {Harmonic generation in {CO$_2$} laser
  target interaction},\ }\href@noop {} {\bibfield  {journal} {\bibinfo
  {journal} {Appl. Phys. Lett.}\ }\textbf {\bibinfo {volume} {31}},\ \bibinfo
  {pages} {172} (\bibinfo {year} {1977})}\BibitemShut {NoStop}%
\bibitem [{\citenamefont {McPherson}\ \emph {et~al.}(1987)\citenamefont
  {McPherson}, \citenamefont {Gibson}, \citenamefont {Jara}, \citenamefont
  {Johann}, \citenamefont {Luk}, \citenamefont {McIntyre}, \citenamefont
  {Boyer},\ and\ \citenamefont {Rhodes}}]{Mcpherson1987studies}%
  \BibitemOpen
  \bibfield  {author} {\bibinfo {author} {\bibfnamefont {A.}~\bibnamefont
  {McPherson}}, \bibinfo {author} {\bibfnamefont {G.}~\bibnamefont {Gibson}},
  \bibinfo {author} {\bibfnamefont {H.}~\bibnamefont {Jara}}, \bibinfo {author}
  {\bibfnamefont {U.}~\bibnamefont {Johann}}, \bibinfo {author} {\bibfnamefont
  {T.~S.}\ \bibnamefont {Luk}}, \bibinfo {author} {\bibfnamefont
  {I.}~\bibnamefont {McIntyre}}, \bibinfo {author} {\bibfnamefont
  {K.}~\bibnamefont {Boyer}}, \ and\ \bibinfo {author} {\bibfnamefont {C.~K.}\
  \bibnamefont {Rhodes}},\ }\bibfield  {title} {\bibinfo {title} {Studies of
  multiphoton production of vacuum-ultraviolet radiation in the rare gases},\
  }\href@noop {} {\bibfield  {journal} {\bibinfo  {journal} {J. Opt. Soc. Am.
  B}\ }\textbf {\bibinfo {volume} {4}},\ \bibinfo {pages} {595} (\bibinfo
  {year} {1987})}\BibitemShut {NoStop}%
\bibitem [{\citenamefont {Ferray}\ \emph {et~al.}(1988)\citenamefont {Ferray},
  \citenamefont {L'Huillier}, \citenamefont {Li}, \citenamefont {Lompre},
  \citenamefont {Mainfray},\ and\ \citenamefont {Manus}}]{Ferray1988multiple}%
  \BibitemOpen
  \bibfield  {author} {\bibinfo {author} {\bibfnamefont {M.}~\bibnamefont
  {Ferray}}, \bibinfo {author} {\bibfnamefont {A.}~\bibnamefont {L'Huillier}},
  \bibinfo {author} {\bibfnamefont {X.}~\bibnamefont {Li}}, \bibinfo {author}
  {\bibfnamefont {L.}~\bibnamefont {Lompre}}, \bibinfo {author} {\bibfnamefont
  {G.}~\bibnamefont {Mainfray}}, \ and\ \bibinfo {author} {\bibfnamefont
  {C.}~\bibnamefont {Manus}},\ }\bibfield  {title} {\bibinfo {title}
  {Multiple-harmonic conversion of 1064 nm radiation in rare gases},\
  }\href@noop {} {\bibfield  {journal} {\bibinfo  {journal} {J. Phys. B}\
  }\textbf {\bibinfo {volume} {21}},\ \bibinfo {pages} {L31} (\bibinfo {year}
  {1988})}\BibitemShut {NoStop}%
\bibitem [{\citenamefont {Gibbon}(1996)}]{Gibbon1996harmonic}%
  \BibitemOpen
  \bibfield  {author} {\bibinfo {author} {\bibfnamefont {P.}~\bibnamefont
  {Gibbon}},\ }\bibfield  {title} {\bibinfo {title} {Harmonic generation by
  femtosecond laser-solid interaction: A coherent {``water-window''} light
  source{?}}\ }\href@noop {} {\bibfield  {journal} {\bibinfo  {journal} {Phys.
  Rev. Lett.}\ }\textbf {\bibinfo {volume} {76}},\ \bibinfo {pages} {50}
  (\bibinfo {year} {1996})}\BibitemShut {NoStop}%
\bibitem [{\citenamefont {Lichters}\ \emph {et~al.}(1996)\citenamefont
  {Lichters}, \citenamefont {Meyer-ter Vehn},\ and\ \citenamefont
  {Pukhov}}]{Lichters1996short}%
  \BibitemOpen
  \bibfield  {author} {\bibinfo {author} {\bibfnamefont {R.}~\bibnamefont
  {Lichters}}, \bibinfo {author} {\bibfnamefont {J.}~\bibnamefont {Meyer-ter
  Vehn}}, \ and\ \bibinfo {author} {\bibfnamefont {A.}~\bibnamefont {Pukhov}},\
  }\bibfield  {title} {\bibinfo {title} {Short-pulse laser harmonics from
  oscillating plasma surfaces driven at relativistic intensity},\ }\href@noop
  {} {\bibfield  {journal} {\bibinfo  {journal} {Phys. Plasmas}\ }\textbf
  {\bibinfo {volume} {3}},\ \bibinfo {pages} {3425} (\bibinfo {year}
  {1996})}\BibitemShut {NoStop}%
\bibitem [{\citenamefont {Chang}\ \emph {et~al.}(1997)\citenamefont {Chang},
  \citenamefont {Rundquist}, \citenamefont {Wang}, \citenamefont {Murnane},\
  and\ \citenamefont {Kapteyn}}]{Chang1997generation}%
  \BibitemOpen
  \bibfield  {author} {\bibinfo {author} {\bibfnamefont {Z.}~\bibnamefont
  {Chang}}, \bibinfo {author} {\bibfnamefont {A.}~\bibnamefont {Rundquist}},
  \bibinfo {author} {\bibfnamefont {H.}~\bibnamefont {Wang}}, \bibinfo {author}
  {\bibfnamefont {M.~M.}\ \bibnamefont {Murnane}}, \ and\ \bibinfo {author}
  {\bibfnamefont {H.~C.}\ \bibnamefont {Kapteyn}},\ }\bibfield  {title}
  {\bibinfo {title} {Generation of coherent soft {X} rays at 2.7 nm using high
  harmonics},\ }\href@noop {} {\bibfield  {journal} {\bibinfo  {journal} {Phys.
  Rev. Lett.}\ }\textbf {\bibinfo {volume} {79}},\ \bibinfo {pages} {2967}
  (\bibinfo {year} {1997})}\BibitemShut {NoStop}%
\bibitem [{\citenamefont {Hentschel}\ \emph {et~al.}(2001)\citenamefont
  {Hentschel}, \citenamefont {Kienberger}, \citenamefont {Spielmann},
  \citenamefont {Reider}, \citenamefont {Milosevic}, \citenamefont {Brabec},
  \citenamefont {Corkum}, \citenamefont {Heinzmann}, \citenamefont {Drescher},\
  and\ \citenamefont {Krausz}}]{Hentschel2001attosecond}%
  \BibitemOpen
  \bibfield  {author} {\bibinfo {author} {\bibfnamefont {M.}~\bibnamefont
  {Hentschel}}, \bibinfo {author} {\bibfnamefont {R.}~\bibnamefont
  {Kienberger}}, \bibinfo {author} {\bibfnamefont {C.}~\bibnamefont
  {Spielmann}}, \bibinfo {author} {\bibfnamefont {G.~A.}\ \bibnamefont
  {Reider}}, \bibinfo {author} {\bibfnamefont {N.}~\bibnamefont {Milosevic}},
  \bibinfo {author} {\bibfnamefont {T.}~\bibnamefont {Brabec}}, \bibinfo
  {author} {\bibfnamefont {P.}~\bibnamefont {Corkum}}, \bibinfo {author}
  {\bibfnamefont {U.}~\bibnamefont {Heinzmann}}, \bibinfo {author}
  {\bibfnamefont {M.}~\bibnamefont {Drescher}}, \ and\ \bibinfo {author}
  {\bibfnamefont {F.}~\bibnamefont {Krausz}},\ }\bibfield  {title} {\bibinfo
  {title} {Attosecond metrology},\ }\href@noop {} {\bibfield  {journal}
  {\bibinfo  {journal} {Nature}\ }\textbf {\bibinfo {volume} {414}},\ \bibinfo
  {pages} {509} (\bibinfo {year} {2001})}\BibitemShut {NoStop}%
\bibitem [{\citenamefont {Qu{\'e}r{\'e}}\ \emph {et~al.}(2006)\citenamefont
  {Qu{\'e}r{\'e}}, \citenamefont {Thaury}, \citenamefont {Monot}, \citenamefont
  {Dobosz}, \citenamefont {Martin}, \citenamefont {Geindre},\ and\
  \citenamefont {Audebert}}]{Quere2006coherent}%
  \BibitemOpen
  \bibfield  {author} {\bibinfo {author} {\bibfnamefont {F.}~\bibnamefont
  {Qu{\'e}r{\'e}}}, \bibinfo {author} {\bibfnamefont {C.}~\bibnamefont
  {Thaury}}, \bibinfo {author} {\bibfnamefont {P.}~\bibnamefont {Monot}},
  \bibinfo {author} {\bibfnamefont {S.}~\bibnamefont {Dobosz}}, \bibinfo
  {author} {\bibfnamefont {P.}~\bibnamefont {Martin}}, \bibinfo {author}
  {\bibfnamefont {J.-P.}\ \bibnamefont {Geindre}}, \ and\ \bibinfo {author}
  {\bibfnamefont {P.}~\bibnamefont {Audebert}},\ }\bibfield  {title} {\bibinfo
  {title} {Coherent wake emission of high-order harmonics from overdense
  plasmas},\ }\href@noop {} {\bibfield  {journal} {\bibinfo  {journal} {Phys.
  Rev. Lett.}\ }\textbf {\bibinfo {volume} {96}},\ \bibinfo {pages} {125004}
  (\bibinfo {year} {2006})}\BibitemShut {NoStop}%
\bibitem [{\citenamefont {Dromey}\ \emph {et~al.}(2007)\citenamefont {Dromey},
  \citenamefont {Kar}, \citenamefont {Bellei}, \citenamefont {Carroll},
  \citenamefont {Clarke}, \citenamefont {Green}, \citenamefont {Kneip},
  \citenamefont {Markey}, \citenamefont {Nagel}, \citenamefont {Simpson} \emph
  {et~al.}}]{Dromey2007bright}%
  \BibitemOpen
  \bibfield  {author} {\bibinfo {author} {\bibfnamefont {B.}~\bibnamefont
  {Dromey}}, \bibinfo {author} {\bibfnamefont {S.}~\bibnamefont {Kar}},
  \bibinfo {author} {\bibfnamefont {C.}~\bibnamefont {Bellei}}, \bibinfo
  {author} {\bibfnamefont {D.}~\bibnamefont {Carroll}}, \bibinfo {author}
  {\bibfnamefont {R.}~\bibnamefont {Clarke}}, \bibinfo {author} {\bibfnamefont
  {J.}~\bibnamefont {Green}}, \bibinfo {author} {\bibfnamefont
  {S.}~\bibnamefont {Kneip}}, \bibinfo {author} {\bibfnamefont
  {K.}~\bibnamefont {Markey}}, \bibinfo {author} {\bibfnamefont
  {S.}~\bibnamefont {Nagel}}, \bibinfo {author} {\bibfnamefont
  {P.}~\bibnamefont {Simpson}},  \emph {et~al.},\ }\bibfield  {title} {\bibinfo
  {title} {Bright multi-kev harmonic generation from relativistically
  oscillating plasma surfaces},\ }\href@noop {} {\bibfield  {journal} {\bibinfo
   {journal} {Phys. Rev. Lett.}\ }\textbf {\bibinfo {volume} {99}},\ \bibinfo
  {pages} {085001} (\bibinfo {year} {2007})}\BibitemShut {NoStop}%
\bibitem [{\citenamefont {Popmintchev}\ \emph {et~al.}(2015)\citenamefont
  {Popmintchev}, \citenamefont {Hern{\'a}ndez-Garc{\'\i}a}, \citenamefont
  {Dollar}, \citenamefont {Mancuso}, \citenamefont {P{\'e}rez-Hern{\'a}ndez},
  \citenamefont {Chen}, \citenamefont {Hankla}, \citenamefont {Gao},
  \citenamefont {Shim}, \citenamefont {Gaeta} \emph
  {et~al.}}]{Popmintchev2015ultraviolet}%
  \BibitemOpen
  \bibfield  {author} {\bibinfo {author} {\bibfnamefont {D.}~\bibnamefont
  {Popmintchev}}, \bibinfo {author} {\bibfnamefont {C.}~\bibnamefont
  {Hern{\'a}ndez-Garc{\'\i}a}}, \bibinfo {author} {\bibfnamefont
  {F.}~\bibnamefont {Dollar}}, \bibinfo {author} {\bibfnamefont
  {C.}~\bibnamefont {Mancuso}}, \bibinfo {author} {\bibfnamefont {J.~A.}\
  \bibnamefont {P{\'e}rez-Hern{\'a}ndez}}, \bibinfo {author} {\bibfnamefont
  {M.-C.}\ \bibnamefont {Chen}}, \bibinfo {author} {\bibfnamefont
  {A.}~\bibnamefont {Hankla}}, \bibinfo {author} {\bibfnamefont
  {X.}~\bibnamefont {Gao}}, \bibinfo {author} {\bibfnamefont {B.}~\bibnamefont
  {Shim}}, \bibinfo {author} {\bibfnamefont {A.~L.}\ \bibnamefont {Gaeta}},
  \emph {et~al.},\ }\bibfield  {title} {\bibinfo {title} {Ultraviolet surprise:
  Efficient soft x-ray high-harmonic generation in multiply ionized plasmas},\
  }\href@noop {} {\bibfield  {journal} {\bibinfo  {journal} {Science}\ }\textbf
  {\bibinfo {volume} {350}},\ \bibinfo {pages} {1225} (\bibinfo {year}
  {2015})}\BibitemShut {NoStop}%
\bibitem [{\citenamefont {Edwards}\ and\ \citenamefont
  {Mikhailova}(2016)}]{Edwards2016waveform}%
  \BibitemOpen
  \bibfield  {author} {\bibinfo {author} {\bibfnamefont {M.~R.}\ \bibnamefont
  {Edwards}}\ and\ \bibinfo {author} {\bibfnamefont {J.~M.}\ \bibnamefont
  {Mikhailova}},\ }\bibfield  {title} {\bibinfo {title} {Waveform-controlled
  relativistic high-order-harmonic generation},\ }\href@noop {} {\bibfield
  {journal} {\bibinfo  {journal} {Phys. Rev. Lett.}\ }\textbf {\bibinfo
  {volume} {117}},\ \bibinfo {pages} {125001} (\bibinfo {year}
  {2016})}\BibitemShut {NoStop}%
\bibitem [{\citenamefont {Malkin}\ and\ \citenamefont
  {Fisch}(2007)}]{Malkin2007relic}%
  \BibitemOpen
  \bibfield  {author} {\bibinfo {author} {\bibfnamefont {V.~M.}\ \bibnamefont
  {Malkin}}\ and\ \bibinfo {author} {\bibfnamefont {N.~J.}\ \bibnamefont
  {Fisch}},\ }\bibfield  {title} {\bibinfo {title} {{Relic crystal-lattice
  effects on Raman compression of powerful x-ray pulses in plasmas}},\
  }\href@noop {} {\bibfield  {journal} {\bibinfo  {journal} {Phys. Rev. Lett.}\
  }\textbf {\bibinfo {volume} {99}},\ \bibinfo {pages} {205001} (\bibinfo
  {year} {2007})}\BibitemShut {NoStop}%
\bibitem [{\citenamefont {Malkin}\ and\ \citenamefont
  {Fisch}(2009)}]{Malkin2009quasitransient}%
  \BibitemOpen
  \bibfield  {author} {\bibinfo {author} {\bibfnamefont {V.~M.}\ \bibnamefont
  {Malkin}}\ and\ \bibinfo {author} {\bibfnamefont {N.~J.}\ \bibnamefont
  {Fisch}},\ }\bibfield  {title} {\bibinfo {title} {{Quasitransient regimes of
  backward Raman amplification of intense x-ray pulses}},\ }\href@noop {}
  {\bibfield  {journal} {\bibinfo  {journal} {Phys. Rev. E}\ }\textbf {\bibinfo
  {volume} {80}},\ \bibinfo {pages} {046409} (\bibinfo {year}
  {2009})}\BibitemShut {NoStop}%
\bibitem [{\citenamefont {Malkin}\ and\ \citenamefont
  {Fisch}(2010)}]{Malkin2010quasitransient}%
  \BibitemOpen
  \bibfield  {author} {\bibinfo {author} {\bibfnamefont {V.}~\bibnamefont
  {Malkin}}\ and\ \bibinfo {author} {\bibfnamefont {N.}~\bibnamefont {Fisch}},\
  }\bibfield  {title} {\bibinfo {title} {{Quasitransient backward Raman
  amplification of powerful laser pulses in dense plasmas with multicharged
  ions}},\ }\href@noop {} {\bibfield  {journal} {\bibinfo  {journal} {Phys.
  Plasmas}\ }\textbf {\bibinfo {volume} {17}},\ \bibinfo {pages} {073109}
  (\bibinfo {year} {2010})}\BibitemShut {NoStop}%
\bibitem [{\citenamefont {Sadler}\ \emph {et~al.}(2015)\citenamefont {Sadler},
  \citenamefont {Nathvani}, \citenamefont {Ole{\'s}kiewicz}, \citenamefont
  {Ceurvorst}, \citenamefont {Ratan}, \citenamefont {Kasim}, \citenamefont
  {Trines}, \citenamefont {Bingham},\ and\ \citenamefont
  {Norreys}}]{Sadler2015compression}%
  \BibitemOpen
  \bibfield  {author} {\bibinfo {author} {\bibfnamefont {J.~D.}\ \bibnamefont
  {Sadler}}, \bibinfo {author} {\bibfnamefont {R.}~\bibnamefont {Nathvani}},
  \bibinfo {author} {\bibfnamefont {P.}~\bibnamefont {Ole{\'s}kiewicz}},
  \bibinfo {author} {\bibfnamefont {L.~A.}\ \bibnamefont {Ceurvorst}}, \bibinfo
  {author} {\bibfnamefont {N.}~\bibnamefont {Ratan}}, \bibinfo {author}
  {\bibfnamefont {M.~F.}\ \bibnamefont {Kasim}}, \bibinfo {author}
  {\bibfnamefont {R.~M. G.~M.}\ \bibnamefont {Trines}}, \bibinfo {author}
  {\bibfnamefont {R.}~\bibnamefont {Bingham}}, \ and\ \bibinfo {author}
  {\bibfnamefont {P.~A.}\ \bibnamefont {Norreys}},\ }\bibfield  {title}
  {\bibinfo {title} {Compression of x-ray free electron laser pulses to
  attosecond duration},\ }\href@noop {} {\bibfield  {journal} {\bibinfo
  {journal} {Sci. Rep.}\ }\textbf {\bibinfo {volume} {5}} (\bibinfo {year}
  {2015})}\BibitemShut {NoStop}%
\bibitem [{\citenamefont {Shi}\ \emph {et~al.}(2017)\citenamefont {Shi},
  \citenamefont {Qin},\ and\ \citenamefont {Fisch}}]{Shi2017Laser}%
  \BibitemOpen
  \bibfield  {author} {\bibinfo {author} {\bibfnamefont {Y.}~\bibnamefont
  {Shi}}, \bibinfo {author} {\bibfnamefont {H.}~\bibnamefont {Qin}}, \ and\
  \bibinfo {author} {\bibfnamefont {N.~J.}\ \bibnamefont {Fisch}},\ }\bibfield
  {title} {\bibinfo {title} {Laser pulse compression using magnetized
  plasmas},\ }\href@noop {} {\bibfield  {journal} {\bibinfo  {journal} {Phys.
  Rev. E}\ }\textbf {\bibinfo {volume} {95}} (\bibinfo {year}
  {2017})}\BibitemShut {NoStop}%
\bibitem [{\citenamefont {Malkin}\ \emph {et~al.}(1999)\citenamefont {Malkin},
  \citenamefont {Shvets},\ and\ \citenamefont {Fisch}}]{Malkin1999}%
  \BibitemOpen
  \bibfield  {author} {\bibinfo {author} {\bibfnamefont {V.~M.}\ \bibnamefont
  {Malkin}}, \bibinfo {author} {\bibfnamefont {G.}~\bibnamefont {Shvets}}, \
  and\ \bibinfo {author} {\bibfnamefont {N.~J.}\ \bibnamefont {Fisch}},\
  }\bibfield  {title} {\bibinfo {title} {{Fast compression of laser beams to
  highly overcritical powers}},\ }\href@noop {} {\bibfield  {journal} {\bibinfo
   {journal} {Phys. Rev. Lett.}\ }\textbf {\bibinfo {volume} {82}},\ \bibinfo
  {pages} {4448} (\bibinfo {year} {1999})}\BibitemShut {NoStop}%
\bibitem [{\citenamefont {Edwards}\ \emph {et~al.}(2016)\citenamefont
  {Edwards}, \citenamefont {Jia}, \citenamefont {Mikhailova},\ and\
  \citenamefont {Fisch}}]{Edwards2016short}%
  \BibitemOpen
  \bibfield  {author} {\bibinfo {author} {\bibfnamefont {M.~R.}\ \bibnamefont
  {Edwards}}, \bibinfo {author} {\bibfnamefont {Q.}~\bibnamefont {Jia}},
  \bibinfo {author} {\bibfnamefont {J.~M.}\ \bibnamefont {Mikhailova}}, \ and\
  \bibinfo {author} {\bibfnamefont {N.~J.}\ \bibnamefont {Fisch}},\ }\bibfield
  {title} {\bibinfo {title} {{Short-pulse amplification by strongly-coupled
  stimulated Brillouin scattering}},\ }\href@noop {} {\bibfield  {journal}
  {\bibinfo  {journal} {Phys. Plasmas}\ }\textbf {\bibinfo {volume} {23}},\
  \bibinfo {pages} {083122} (\bibinfo {year} {2016})}\BibitemShut {NoStop}%
\bibitem [{\citenamefont {Malkin}\ \emph {et~al.}(2000)\citenamefont {Malkin},
  \citenamefont {Shvets},\ and\ \citenamefont {Fisch}}]{Malkin2000}%
  \BibitemOpen
  \bibfield  {author} {\bibinfo {author} {\bibfnamefont {V.}~\bibnamefont
  {Malkin}}, \bibinfo {author} {\bibfnamefont {G.}~\bibnamefont {Shvets}}, \
  and\ \bibinfo {author} {\bibfnamefont {N.}~\bibnamefont {Fisch}},\ }\bibfield
   {title} {\bibinfo {title} {Ultra-powerful compact amplifiers for short laser
  pulses},\ }\href@noop {} {\bibfield  {journal} {\bibinfo  {journal} {Phys.
  Plasmas}\ }\textbf {\bibinfo {volume} {7}},\ \bibinfo {pages} {2232}
  (\bibinfo {year} {2000})}\BibitemShut {NoStop}%
\bibitem [{\citenamefont {Ping}\ \emph {et~al.}(2004)\citenamefont {Ping},
  \citenamefont {Cheng}, \citenamefont {Suckewer}, \citenamefont {Clark},\ and\
  \citenamefont {Fisch}}]{Ping2004}%
  \BibitemOpen
  \bibfield  {author} {\bibinfo {author} {\bibfnamefont {Y.}~\bibnamefont
  {Ping}}, \bibinfo {author} {\bibfnamefont {W.}~\bibnamefont {Cheng}},
  \bibinfo {author} {\bibfnamefont {S.}~\bibnamefont {Suckewer}}, \bibinfo
  {author} {\bibfnamefont {D.~S.}\ \bibnamefont {Clark}}, \ and\ \bibinfo
  {author} {\bibfnamefont {N.~J.}\ \bibnamefont {Fisch}},\ }\bibfield  {title}
  {\bibinfo {title} {{Amplification of ultrashort laser pulses by a resonant
  Raman scheme in a gas-jet plasma}},\ }\href@noop {} {\bibfield  {journal}
  {\bibinfo  {journal} {Phys. Rev. Lett.}\ }\textbf {\bibinfo {volume} {92}},\
  \bibinfo {pages} {175007} (\bibinfo {year} {2004})}\BibitemShut {NoStop}%
\bibitem [{\citenamefont {Cheng}\ \emph {et~al.}(2005)\citenamefont {Cheng},
  \citenamefont {Avitzour}, \citenamefont {Ping}, \citenamefont {Suckewer},
  \citenamefont {Fisch}, \citenamefont {Hur},\ and\ \citenamefont
  {Wurtele}}]{Cheng2005}%
  \BibitemOpen
  \bibfield  {author} {\bibinfo {author} {\bibfnamefont {W.}~\bibnamefont
  {Cheng}}, \bibinfo {author} {\bibfnamefont {Y.}~\bibnamefont {Avitzour}},
  \bibinfo {author} {\bibfnamefont {Y.}~\bibnamefont {Ping}}, \bibinfo {author}
  {\bibfnamefont {S.}~\bibnamefont {Suckewer}}, \bibinfo {author}
  {\bibfnamefont {N.~J.}\ \bibnamefont {Fisch}}, \bibinfo {author}
  {\bibfnamefont {M.~S.}\ \bibnamefont {Hur}}, \ and\ \bibinfo {author}
  {\bibfnamefont {J.~S.}\ \bibnamefont {Wurtele}},\ }\bibfield  {title}
  {\bibinfo {title} {{Reaching the nonlinear regime of Raman amplification of
  ultrashort laser pulses}},\ }\href@noop {} {\bibfield  {journal} {\bibinfo
  {journal} {Phys. Rev. Lett.}\ }\textbf {\bibinfo {volume} {94}},\ \bibinfo
  {pages} {045003} (\bibinfo {year} {2005})}\BibitemShut {NoStop}%
\bibitem [{\citenamefont {Ren}\ \emph {et~al.}(2008)\citenamefont {Ren},
  \citenamefont {Li}, \citenamefont {Morozov}, \citenamefont {Suckewer},
  \citenamefont {Yampolsky}, \citenamefont {Malkin},\ and\ \citenamefont
  {Fisch}}]{Ren2008}%
  \BibitemOpen
  \bibfield  {author} {\bibinfo {author} {\bibfnamefont {J.}~\bibnamefont
  {Ren}}, \bibinfo {author} {\bibfnamefont {S.}~\bibnamefont {Li}}, \bibinfo
  {author} {\bibfnamefont {A.}~\bibnamefont {Morozov}}, \bibinfo {author}
  {\bibfnamefont {S.}~\bibnamefont {Suckewer}}, \bibinfo {author}
  {\bibfnamefont {N.}~\bibnamefont {Yampolsky}}, \bibinfo {author}
  {\bibfnamefont {V.}~\bibnamefont {Malkin}}, \ and\ \bibinfo {author}
  {\bibfnamefont {N.}~\bibnamefont {Fisch}},\ }\bibfield  {title} {\bibinfo
  {title} {{A compact double-pass Raman backscattering amplifier/compressor}},\
  }\href@noop {} {\bibfield  {journal} {\bibinfo  {journal} {Phys. Plasmas}\
  }\textbf {\bibinfo {volume} {15}},\ \bibinfo {pages} {056702} (\bibinfo
  {year} {2008})}\BibitemShut {NoStop}%
\bibitem [{\citenamefont {Yampolsky}\ \emph {et~al.}(2008)\citenamefont
  {Yampolsky}, \citenamefont {Fisch}, \citenamefont {Malkin}, \citenamefont
  {Valeo}, \citenamefont {Lindberg}, \citenamefont {Wurtele}, \citenamefont
  {Ren}, \citenamefont {Li}, \citenamefont {Morozov},\ and\ \citenamefont
  {Suckewer}}]{Yampolsky2008}%
  \BibitemOpen
  \bibfield  {author} {\bibinfo {author} {\bibfnamefont {N.}~\bibnamefont
  {Yampolsky}}, \bibinfo {author} {\bibfnamefont {N.}~\bibnamefont {Fisch}},
  \bibinfo {author} {\bibfnamefont {V.}~\bibnamefont {Malkin}}, \bibinfo
  {author} {\bibfnamefont {E.}~\bibnamefont {Valeo}}, \bibinfo {author}
  {\bibfnamefont {R.}~\bibnamefont {Lindberg}}, \bibinfo {author}
  {\bibfnamefont {J.}~\bibnamefont {Wurtele}}, \bibinfo {author} {\bibfnamefont
  {J.}~\bibnamefont {Ren}}, \bibinfo {author} {\bibfnamefont {S.}~\bibnamefont
  {Li}}, \bibinfo {author} {\bibfnamefont {A.}~\bibnamefont {Morozov}}, \ and\
  \bibinfo {author} {\bibfnamefont {S.}~\bibnamefont {Suckewer}},\ }\bibfield
  {title} {\bibinfo {title} {{Demonstration of detuning and wavebreaking
  effects on Raman amplification efficiency in plasma}},\ }\href@noop {}
  {\bibfield  {journal} {\bibinfo  {journal} {Phys. Plasmas}\ }\textbf
  {\bibinfo {volume} {15}},\ \bibinfo {pages} {113104} (\bibinfo {year}
  {2008})}\BibitemShut {NoStop}%
\bibitem [{\citenamefont {Ping}\ \emph {et~al.}(2009)\citenamefont {Ping},
  \citenamefont {Kirkwood}, \citenamefont {Wang}, \citenamefont {Clark},
  \citenamefont {Wilks}, \citenamefont {Meezan}, \citenamefont {Berger},
  \citenamefont {Wurtele}, \citenamefont {Fisch}, \citenamefont {Malkin} \emph
  {et~al.}}]{Ping2009}%
  \BibitemOpen
  \bibfield  {author} {\bibinfo {author} {\bibfnamefont {Y.}~\bibnamefont
  {Ping}}, \bibinfo {author} {\bibfnamefont {R.}~\bibnamefont {Kirkwood}},
  \bibinfo {author} {\bibfnamefont {T.-L.}\ \bibnamefont {Wang}}, \bibinfo
  {author} {\bibfnamefont {D.}~\bibnamefont {Clark}}, \bibinfo {author}
  {\bibfnamefont {S.}~\bibnamefont {Wilks}}, \bibinfo {author} {\bibfnamefont
  {N.}~\bibnamefont {Meezan}}, \bibinfo {author} {\bibfnamefont
  {R.}~\bibnamefont {Berger}}, \bibinfo {author} {\bibfnamefont
  {J.}~\bibnamefont {Wurtele}}, \bibinfo {author} {\bibfnamefont
  {N.}~\bibnamefont {Fisch}}, \bibinfo {author} {\bibfnamefont
  {V.}~\bibnamefont {Malkin}},  \emph {et~al.},\ }\bibfield  {title} {\bibinfo
  {title} {{Development of a nanosecond-laser-pumped Raman amplifier for short
  laser pulses in plasma}},\ }\href@noop {} {\bibfield  {journal} {\bibinfo
  {journal} {Phys. Plasmas}\ }\textbf {\bibinfo {volume} {16}},\ \bibinfo
  {pages} {123113} (\bibinfo {year} {2009})}\BibitemShut {NoStop}%
\bibitem [{\citenamefont {Vieux}\ \emph {et~al.}(2011)\citenamefont {Vieux},
  \citenamefont {Lyachev}, \citenamefont {Yang}, \citenamefont {Ersfeld},
  \citenamefont {Farmer}, \citenamefont {Brunetti}, \citenamefont {Issac},
  \citenamefont {Raj}, \citenamefont {Welsh}, \citenamefont {Wiggins} \emph
  {et~al.}}]{Vieux2011}%
  \BibitemOpen
  \bibfield  {author} {\bibinfo {author} {\bibfnamefont {G.}~\bibnamefont
  {Vieux}}, \bibinfo {author} {\bibfnamefont {A.}~\bibnamefont {Lyachev}},
  \bibinfo {author} {\bibfnamefont {X.}~\bibnamefont {Yang}}, \bibinfo {author}
  {\bibfnamefont {B.}~\bibnamefont {Ersfeld}}, \bibinfo {author} {\bibfnamefont
  {J.}~\bibnamefont {Farmer}}, \bibinfo {author} {\bibfnamefont
  {E.}~\bibnamefont {Brunetti}}, \bibinfo {author} {\bibfnamefont
  {R.}~\bibnamefont {Issac}}, \bibinfo {author} {\bibfnamefont
  {G.}~\bibnamefont {Raj}}, \bibinfo {author} {\bibfnamefont {G.}~\bibnamefont
  {Welsh}}, \bibinfo {author} {\bibfnamefont {S.}~\bibnamefont {Wiggins}},
  \emph {et~al.},\ }\bibfield  {title} {\bibinfo {title} {{Chirped pulse Raman
  amplification in plasma}},\ }\href@noop {} {\bibfield  {journal} {\bibinfo
  {journal} {New J. Phys.}\ }\textbf {\bibinfo {volume} {13}},\ \bibinfo
  {pages} {063042} (\bibinfo {year} {2011})}\BibitemShut {NoStop}%
\bibitem [{\citenamefont {Trines}\ \emph {et~al.}(2011)\citenamefont {Trines},
  \citenamefont {Fi{\'u}za}, \citenamefont {Bingham}, \citenamefont {Fonseca},
  \citenamefont {Silva}, \citenamefont {Cairns},\ and\ \citenamefont
  {Norreys}}]{Trines2011production}%
  \BibitemOpen
  \bibfield  {author} {\bibinfo {author} {\bibfnamefont {R.~M. G.~M.}\
  \bibnamefont {Trines}}, \bibinfo {author} {\bibfnamefont {F.}~\bibnamefont
  {Fi{\'u}za}}, \bibinfo {author} {\bibfnamefont {R.}~\bibnamefont {Bingham}},
  \bibinfo {author} {\bibfnamefont {R.~A.}\ \bibnamefont {Fonseca}}, \bibinfo
  {author} {\bibfnamefont {L.~O.}\ \bibnamefont {Silva}}, \bibinfo {author}
  {\bibfnamefont {R.~A.}\ \bibnamefont {Cairns}}, \ and\ \bibinfo {author}
  {\bibfnamefont {P.~A.}\ \bibnamefont {Norreys}},\ }\bibfield  {title}
  {\bibinfo {title} {{Production of picosecond, kilojoule, and petawatt laser
  pulses via Raman amplification of nanosecond pulses}},\ }\href@noop {}
  {\bibfield  {journal} {\bibinfo  {journal} {Phys. Rev. Lett.}\ }\textbf
  {\bibinfo {volume} {107}},\ \bibinfo {pages} {105002} (\bibinfo {year}
  {2011})}\BibitemShut {NoStop}%
\bibitem [{\citenamefont {Turnbull}\ \emph {et~al.}(2012)\citenamefont
  {Turnbull}, \citenamefont {Li}, \citenamefont {Morozov},\ and\ \citenamefont
  {Suckewer}}]{Turnbull2012}%
  \BibitemOpen
  \bibfield  {author} {\bibinfo {author} {\bibfnamefont {D.}~\bibnamefont
  {Turnbull}}, \bibinfo {author} {\bibfnamefont {S.}~\bibnamefont {Li}},
  \bibinfo {author} {\bibfnamefont {A.}~\bibnamefont {Morozov}}, \ and\
  \bibinfo {author} {\bibfnamefont {S.}~\bibnamefont {Suckewer}},\ }\bibfield
  {title} {\bibinfo {title} {{Possible origins of a time-resolved frequency
  shift in Raman plasma amplifiers}},\ }\href@noop {} {\bibfield  {journal}
  {\bibinfo  {journal} {Phys. Plasmas}\ }\textbf {\bibinfo {volume} {19}},\
  \bibinfo {pages} {073103} (\bibinfo {year} {2012})}\BibitemShut {NoStop}%
\bibitem [{\citenamefont {Depierreux}\ \emph {et~al.}(2014)\citenamefont
  {Depierreux}, \citenamefont {Yahia}, \citenamefont {Goyon}, \citenamefont
  {Loisel}, \citenamefont {Masson-Laborde}, \citenamefont {Borisenko},
  \citenamefont {Orekhov}, \citenamefont {Rosmej}, \citenamefont {Rienecker},\
  and\ \citenamefont {Labaune}}]{Depierreux2014}%
  \BibitemOpen
  \bibfield  {author} {\bibinfo {author} {\bibfnamefont {S.}~\bibnamefont
  {Depierreux}}, \bibinfo {author} {\bibfnamefont {V.}~\bibnamefont {Yahia}},
  \bibinfo {author} {\bibfnamefont {C.}~\bibnamefont {Goyon}}, \bibinfo
  {author} {\bibfnamefont {G.}~\bibnamefont {Loisel}}, \bibinfo {author}
  {\bibfnamefont {P.-E.}\ \bibnamefont {Masson-Laborde}}, \bibinfo {author}
  {\bibfnamefont {N.}~\bibnamefont {Borisenko}}, \bibinfo {author}
  {\bibfnamefont {A.}~\bibnamefont {Orekhov}}, \bibinfo {author} {\bibfnamefont
  {O.}~\bibnamefont {Rosmej}}, \bibinfo {author} {\bibfnamefont
  {T.}~\bibnamefont {Rienecker}}, \ and\ \bibinfo {author} {\bibfnamefont
  {C.}~\bibnamefont {Labaune}},\ }\bibfield  {title} {\bibinfo {title} {{Laser
  light triggers increased Raman amplification in the regime of nonlinear
  Landau damping}},\ }\href@noop {} {\bibfield  {journal} {\bibinfo  {journal}
  {Nat. Commun.}\ }\textbf {\bibinfo {volume} {5}} (\bibinfo {year}
  {2014})}\BibitemShut {NoStop}%
\bibitem [{\citenamefont {Toroker}\ \emph {et~al.}(2014)\citenamefont
  {Toroker}, \citenamefont {Malkin},\ and\ \citenamefont
  {Fisch}}]{Toroker2014}%
  \BibitemOpen
  \bibfield  {author} {\bibinfo {author} {\bibfnamefont {Z.}~\bibnamefont
  {Toroker}}, \bibinfo {author} {\bibfnamefont {V.}~\bibnamefont {Malkin}}, \
  and\ \bibinfo {author} {\bibfnamefont {N.}~\bibnamefont {Fisch}},\ }\bibfield
   {title} {\bibinfo {title} {{Backward Raman amplification in the Langmuir
  wavebreaking regime}},\ }\href@noop {} {\bibfield  {journal} {\bibinfo
  {journal} {Phys. Plasmas}\ }\textbf {\bibinfo {volume} {21}},\ \bibinfo
  {pages} {113110} (\bibinfo {year} {2014})}\BibitemShut {NoStop}%
\bibitem [{\citenamefont {Lehmann}\ and\ \citenamefont
  {Spatschek}(2014)}]{Lehmann2014}%
  \BibitemOpen
  \bibfield  {author} {\bibinfo {author} {\bibfnamefont {G.}~\bibnamefont
  {Lehmann}}\ and\ \bibinfo {author} {\bibfnamefont {K.}~\bibnamefont
  {Spatschek}},\ }\bibfield  {title} {\bibinfo {title} {{Non-filamentated
  ultra-intense and ultra-short pulse fronts in three-dimensional Raman seed
  amplification}},\ }\href@noop {} {\bibfield  {journal} {\bibinfo  {journal}
  {Phys. Plasmas}\ }\textbf {\bibinfo {volume} {21}},\ \bibinfo {pages}
  {053101} (\bibinfo {year} {2014})}\BibitemShut {NoStop}%
\bibitem [{\citenamefont {Edwards}\ \emph {et~al.}(2015)\citenamefont
  {Edwards}, \citenamefont {Toroker}, \citenamefont {Mikhailova},\ and\
  \citenamefont {Fisch}}]{Edwards2015}%
  \BibitemOpen
  \bibfield  {author} {\bibinfo {author} {\bibfnamefont {M.~R.}\ \bibnamefont
  {Edwards}}, \bibinfo {author} {\bibfnamefont {Z.}~\bibnamefont {Toroker}},
  \bibinfo {author} {\bibfnamefont {J.~M.}\ \bibnamefont {Mikhailova}}, \ and\
  \bibinfo {author} {\bibfnamefont {N.~J.}\ \bibnamefont {Fisch}},\ }\bibfield
  {title} {\bibinfo {title} {{The efficiency of Raman amplification in the
  wavebreaking regime}},\ }\href@noop {} {\bibfield  {journal} {\bibinfo
  {journal} {Phys. Plasmas}\ }\textbf {\bibinfo {volume} {22}},\ \bibinfo {eid}
  {074501} (\bibinfo {year} {2015})}\BibitemShut {NoStop}%
\bibitem [{\citenamefont {Milroy}\ \emph {et~al.}(1977)\citenamefont {Milroy},
  \citenamefont {Capjack},\ and\ \citenamefont {James}}]{Milroy1977}%
  \BibitemOpen
  \bibfield  {author} {\bibinfo {author} {\bibfnamefont {R.}~\bibnamefont
  {Milroy}}, \bibinfo {author} {\bibfnamefont {C.}~\bibnamefont {Capjack}}, \
  and\ \bibinfo {author} {\bibfnamefont {C.}~\bibnamefont {James}},\ }\bibfield
   {title} {\bibinfo {title} {A plasma-laser amplifier in the 11-16 $\mu$m
  wavelength range},\ }\href@noop {} {\bibfield  {journal} {\bibinfo  {journal}
  {Plasma Phys.}\ }\textbf {\bibinfo {volume} {19}},\ \bibinfo {pages} {989}
  (\bibinfo {year} {1977})}\BibitemShut {NoStop}%
\bibitem [{\citenamefont {Andreev}\ and\ \citenamefont
  {Sutyagin}(1989)}]{Andreev1989}%
  \BibitemOpen
  \bibfield  {author} {\bibinfo {author} {\bibfnamefont {A.~A.}\ \bibnamefont
  {Andreev}}\ and\ \bibinfo {author} {\bibfnamefont {A.}~\bibnamefont
  {Sutyagin}},\ }\bibfield  {title} {\bibinfo {title} {{Feasibility of optical
  pulse compression by stimulated Brillouin scattering in a plasma}},\
  }\href@noop {} {\bibfield  {journal} {\bibinfo  {journal} {Sov. J. Quant.
  Electron.}\ }\textbf {\bibinfo {volume} {19}},\ \bibinfo {pages} {1579}
  (\bibinfo {year} {1989})}\BibitemShut {NoStop}%
\bibitem [{\citenamefont {Andreev}\ \emph {et~al.}(2006)\citenamefont
  {Andreev}, \citenamefont {Riconda}, \citenamefont {Tikhonchuk},\ and\
  \citenamefont {Weber}}]{Andreev2006}%
  \BibitemOpen
  \bibfield  {author} {\bibinfo {author} {\bibfnamefont {A.}~\bibnamefont
  {Andreev}}, \bibinfo {author} {\bibfnamefont {C.}~\bibnamefont {Riconda}},
  \bibinfo {author} {\bibfnamefont {V.}~\bibnamefont {Tikhonchuk}}, \ and\
  \bibinfo {author} {\bibfnamefont {S.}~\bibnamefont {Weber}},\ }\bibfield
  {title} {\bibinfo {title} {{Short light pulse amplification and compression
  by stimulated Brillouin scattering in plasmas in the strong coupling
  regime}},\ }\href@noop {} {\bibfield  {journal} {\bibinfo  {journal} {Phys.
  Plasmas}\ }\textbf {\bibinfo {volume} {13}},\ \bibinfo {pages} {053110}
  (\bibinfo {year} {2006})}\BibitemShut {NoStop}%
\bibitem [{\citenamefont {Lancia}\ \emph {et~al.}(2010)\citenamefont {Lancia},
  \citenamefont {Marques}, \citenamefont {Nakatsutsumi}, \citenamefont
  {Riconda}, \citenamefont {Weber}, \citenamefont {H{\"u}ller}, \citenamefont
  {Man{\v{c}}i{\'c}}, \citenamefont {Antici}, \citenamefont {Tikhonchuk},
  \citenamefont {H{\'e}ron} \emph {et~al.}}]{Lancia2010}%
  \BibitemOpen
  \bibfield  {author} {\bibinfo {author} {\bibfnamefont {L.}~\bibnamefont
  {Lancia}}, \bibinfo {author} {\bibfnamefont {J.-R.}\ \bibnamefont {Marques}},
  \bibinfo {author} {\bibfnamefont {M.}~\bibnamefont {Nakatsutsumi}}, \bibinfo
  {author} {\bibfnamefont {C.}~\bibnamefont {Riconda}}, \bibinfo {author}
  {\bibfnamefont {S.}~\bibnamefont {Weber}}, \bibinfo {author} {\bibfnamefont
  {S.}~\bibnamefont {H{\"u}ller}}, \bibinfo {author} {\bibfnamefont
  {A.}~\bibnamefont {Man{\v{c}}i{\'c}}}, \bibinfo {author} {\bibfnamefont
  {P.}~\bibnamefont {Antici}}, \bibinfo {author} {\bibfnamefont
  {V.}~\bibnamefont {Tikhonchuk}}, \bibinfo {author} {\bibfnamefont
  {A.}~\bibnamefont {H{\'e}ron}},  \emph {et~al.},\ }\bibfield  {title}
  {\bibinfo {title} {{Experimental evidence of short light pulse amplification
  using strong-coupling stimulated Brillouin scattering in the pump depletion
  regime}},\ }\href@noop {} {\bibfield  {journal} {\bibinfo  {journal} {Phys.
  Rev. Lett.}\ }\textbf {\bibinfo {volume} {104}},\ \bibinfo {pages} {025001}
  (\bibinfo {year} {2010})}\BibitemShut {NoStop}%
\bibitem [{\citenamefont {Lehmann}\ \emph {et~al.}(2012)\citenamefont
  {Lehmann}, \citenamefont {Schluck},\ and\ \citenamefont
  {Spatschek}}]{Lehmann2012}%
  \BibitemOpen
  \bibfield  {author} {\bibinfo {author} {\bibfnamefont {G.}~\bibnamefont
  {Lehmann}}, \bibinfo {author} {\bibfnamefont {F.}~\bibnamefont {Schluck}}, \
  and\ \bibinfo {author} {\bibfnamefont {K.~H.}\ \bibnamefont {Spatschek}},\
  }\bibfield  {title} {\bibinfo {title} {{Regions for Brillouin seed pulse
  growth in relativistic laser-plasma interaction}},\ }\href@noop {} {\bibfield
   {journal} {\bibinfo  {journal} {Phys. Plasmas}\ }\textbf {\bibinfo {volume}
  {19}},\ \bibinfo {pages} {093120} (\bibinfo {year} {2012})}\BibitemShut
  {NoStop}%
\bibitem [{\citenamefont {Lehmann}\ and\ \citenamefont
  {Spatschek}(2013)}]{Lehmann2013}%
  \BibitemOpen
  \bibfield  {author} {\bibinfo {author} {\bibfnamefont {G.}~\bibnamefont
  {Lehmann}}\ and\ \bibinfo {author} {\bibfnamefont {K.}~\bibnamefont
  {Spatschek}},\ }\bibfield  {title} {\bibinfo {title} {{Nonlinear Brillouin
  amplification of finite-duration seeds in the strong coupling regime}},\
  }\href@noop {} {\bibfield  {journal} {\bibinfo  {journal} {Phys. Plasmas}\
  }\textbf {\bibinfo {volume} {20}},\ \bibinfo {pages} {073112} (\bibinfo
  {year} {2013})}\BibitemShut {NoStop}%
\bibitem [{\citenamefont {Weber}\ \emph {et~al.}(2013)\citenamefont {Weber},
  \citenamefont {Riconda}, \citenamefont {Lancia}, \citenamefont {Marqu\`{e}s},
  \citenamefont {Mourou},\ and\ \citenamefont {Fuchs}}]{Weber2013}%
  \BibitemOpen
  \bibfield  {author} {\bibinfo {author} {\bibfnamefont {S.}~\bibnamefont
  {Weber}}, \bibinfo {author} {\bibfnamefont {C.}~\bibnamefont {Riconda}},
  \bibinfo {author} {\bibfnamefont {L.}~\bibnamefont {Lancia}}, \bibinfo
  {author} {\bibfnamefont {J.-R.}\ \bibnamefont {Marqu\`{e}s}}, \bibinfo
  {author} {\bibfnamefont {G.~A.}\ \bibnamefont {Mourou}}, \ and\ \bibinfo
  {author} {\bibfnamefont {J.}~\bibnamefont {Fuchs}},\ }\bibfield  {title}
  {\bibinfo {title} {{Amplification of ultrashort laser pulses by Brillouin
  backscattering in plasmas}},\ }\href@noop {} {\bibfield  {journal} {\bibinfo
  {journal} {Phys. Rev. Lett.}\ }\textbf {\bibinfo {volume} {111}},\ \bibinfo
  {pages} {055004} (\bibinfo {year} {2013})}\BibitemShut {NoStop}%
\bibitem [{\citenamefont {Riconda}\ \emph {et~al.}(2013)\citenamefont
  {Riconda}, \citenamefont {Weber}, \citenamefont {Lancia}, \citenamefont
  {MarqueÌ€s}, \citenamefont {Mourou},\ and\ \citenamefont
  {Fuchs}}]{Riconda2013}%
  \BibitemOpen
  \bibfield  {author} {\bibinfo {author} {\bibfnamefont {C.}~\bibnamefont
  {Riconda}}, \bibinfo {author} {\bibfnamefont {S.}~\bibnamefont {Weber}},
  \bibinfo {author} {\bibfnamefont {L.}~\bibnamefont {Lancia}}, \bibinfo
  {author} {\bibfnamefont {J.-R.}\ \bibnamefont {MarqueÌ€s}}, \bibinfo {author}
  {\bibfnamefont {G.~A.}\ \bibnamefont {Mourou}}, \ and\ \bibinfo {author}
  {\bibfnamefont {J.}~\bibnamefont {Fuchs}},\ }\bibfield  {title} {\bibinfo
  {title} {{Spectral characteristics of ultra-short laser pulses in plasma
  amplifiers}},\ }\href {\doibase 10.1063/1.4818893} {\bibfield  {journal}
  {\bibinfo  {journal} {Phys. Plasmas}\ }\textbf {\bibinfo {volume} {20}},\
  \bibinfo {pages} {083115} (\bibinfo {year} {2013})}\BibitemShut {NoStop}%
\bibitem [{\citenamefont {Guillaume}\ \emph {et~al.}(2014)\citenamefont
  {Guillaume}, \citenamefont {Humphrey}, \citenamefont {Nakamura},
  \citenamefont {Trines}, \citenamefont {Heathcote}, \citenamefont
  {Galimberti}, \citenamefont {Amano}, \citenamefont {Doria}, \citenamefont
  {Hicks}, \citenamefont {Higson} \emph {et~al.}}]{Guillaume2014demonstration}%
  \BibitemOpen
  \bibfield  {author} {\bibinfo {author} {\bibfnamefont {E.}~\bibnamefont
  {Guillaume}}, \bibinfo {author} {\bibfnamefont {K.}~\bibnamefont {Humphrey}},
  \bibinfo {author} {\bibfnamefont {H.}~\bibnamefont {Nakamura}}, \bibinfo
  {author} {\bibfnamefont {R.}~\bibnamefont {Trines}}, \bibinfo {author}
  {\bibfnamefont {R.}~\bibnamefont {Heathcote}}, \bibinfo {author}
  {\bibfnamefont {M.}~\bibnamefont {Galimberti}}, \bibinfo {author}
  {\bibfnamefont {Y.}~\bibnamefont {Amano}}, \bibinfo {author} {\bibfnamefont
  {D.}~\bibnamefont {Doria}}, \bibinfo {author} {\bibfnamefont
  {G.}~\bibnamefont {Hicks}}, \bibinfo {author} {\bibfnamefont
  {E.}~\bibnamefont {Higson}},  \emph {et~al.},\ }\bibfield  {title} {\bibinfo
  {title} {{Demonstration of laser pulse amplification by stimulated Brillouin
  scattering}},\ }\href@noop {} {\bibfield  {journal} {\bibinfo  {journal}
  {High Power Laser Sci. Eng.}\ }\textbf {\bibinfo {volume} {2}},\ \bibinfo
  {pages} {e33} (\bibinfo {year} {2014})}\BibitemShut {NoStop}%
\bibitem [{\citenamefont {Lehmann}\ and\ \citenamefont
  {Spatschek}(2016)}]{Lehmann2016temperature}%
  \BibitemOpen
  \bibfield  {author} {\bibinfo {author} {\bibfnamefont {G.}~\bibnamefont
  {Lehmann}}\ and\ \bibinfo {author} {\bibfnamefont {K.}~\bibnamefont
  {Spatschek}},\ }\bibfield  {title} {\bibinfo {title} {{Temperature dependence
  of seed pulse amplitude and density grating in Brillouin amplification}},\
  }\href@noop {} {\bibfield  {journal} {\bibinfo  {journal} {Phys. Plasmas}\
  }\textbf {\bibinfo {volume} {23}},\ \bibinfo {pages} {023107} (\bibinfo
  {year} {2016})}\BibitemShut {NoStop}%
\bibitem [{\citenamefont {Lancia}\ \emph {et~al.}(2016)\citenamefont {Lancia},
  \citenamefont {Giribono}, \citenamefont {Vassura}, \citenamefont
  {Chiaramello}, \citenamefont {Riconda}, \citenamefont {Weber}, \citenamefont
  {Castan}, \citenamefont {Chatelain}, \citenamefont {Frank}, \citenamefont
  {Gangolf}, \citenamefont {Quinn}, \citenamefont {Fuchs},\ and\ \citenamefont
  {Marqu\`es}}]{Lancia2016signatures}%
  \BibitemOpen
  \bibfield  {author} {\bibinfo {author} {\bibfnamefont {L.}~\bibnamefont
  {Lancia}}, \bibinfo {author} {\bibfnamefont {A.}~\bibnamefont {Giribono}},
  \bibinfo {author} {\bibfnamefont {L.}~\bibnamefont {Vassura}}, \bibinfo
  {author} {\bibfnamefont {M.}~\bibnamefont {Chiaramello}}, \bibinfo {author}
  {\bibfnamefont {C.}~\bibnamefont {Riconda}}, \bibinfo {author} {\bibfnamefont
  {S.}~\bibnamefont {Weber}}, \bibinfo {author} {\bibfnamefont
  {A.}~\bibnamefont {Castan}}, \bibinfo {author} {\bibfnamefont
  {A.}~\bibnamefont {Chatelain}}, \bibinfo {author} {\bibfnamefont
  {A.}~\bibnamefont {Frank}}, \bibinfo {author} {\bibfnamefont
  {T.}~\bibnamefont {Gangolf}}, \bibinfo {author} {\bibfnamefont {M.~N.}\
  \bibnamefont {Quinn}}, \bibinfo {author} {\bibfnamefont {J.}~\bibnamefont
  {Fuchs}}, \ and\ \bibinfo {author} {\bibfnamefont {J.-R.}\ \bibnamefont
  {Marqu\`es}},\ }\bibfield  {title} {\bibinfo {title} {{Signatures of the
  self-similar regime of strongly coupled stimulated Brillouin scattering for
  efficient short laser pulse amplification}},\ }\href@noop {} {\bibfield
  {journal} {\bibinfo  {journal} {Phys. Rev. Lett.}\ }\textbf {\bibinfo
  {volume} {116}},\ \bibinfo {pages} {075001} (\bibinfo {year}
  {2016})}\BibitemShut {NoStop}%
\bibitem [{\citenamefont {Chiaramello}\ \emph {et~al.}(2016)\citenamefont
  {Chiaramello}, \citenamefont {Amiranoff}, \citenamefont {Riconda},\ and\
  \citenamefont {Weber}}]{Chiaramello2016role}%
  \BibitemOpen
  \bibfield  {author} {\bibinfo {author} {\bibfnamefont {M.}~\bibnamefont
  {Chiaramello}}, \bibinfo {author} {\bibfnamefont {F.}~\bibnamefont
  {Amiranoff}}, \bibinfo {author} {\bibfnamefont {C.}~\bibnamefont {Riconda}},
  \ and\ \bibinfo {author} {\bibfnamefont {S.}~\bibnamefont {Weber}},\
  }\bibfield  {title} {\bibinfo {title} {{Role of frequency chirp and energy
  flow directionality in the strong coupling regime of Brillouin-based plasma
  amplification}},\ }\href {\doibase 10.1103/PhysRevLett.117.235003} {\bibfield
   {journal} {\bibinfo  {journal} {Phys. Rev. Lett.}\ }\textbf {\bibinfo
  {volume} {117}},\ \bibinfo {pages} {235003} (\bibinfo {year}
  {2016})}\BibitemShut {NoStop}%
\bibitem [{\citenamefont {Schluck}\ \emph {et~al.}(2016)\citenamefont
  {Schluck}, \citenamefont {Lehmann}, \citenamefont {M{\"u}ller},\ and\
  \citenamefont {Spatschek}}]{Schluck2016dynamical}%
  \BibitemOpen
  \bibfield  {author} {\bibinfo {author} {\bibfnamefont {F.}~\bibnamefont
  {Schluck}}, \bibinfo {author} {\bibfnamefont {G.}~\bibnamefont {Lehmann}},
  \bibinfo {author} {\bibfnamefont {C.}~\bibnamefont {M{\"u}ller}}, \ and\
  \bibinfo {author} {\bibfnamefont {K.}~\bibnamefont {Spatschek}},\ }\bibfield
  {title} {\bibinfo {title} {{Dynamical transition between weak and strong
  coupling in Brillouin laser pulse amplification}},\ }\href@noop {} {\bibfield
   {journal} {\bibinfo  {journal} {Phys. Plasmas}\ }\textbf {\bibinfo {volume}
  {23}},\ \bibinfo {pages} {083105} (\bibinfo {year} {2016})}\BibitemShut
  {NoStop}%
\bibitem [{\citenamefont {Jia}\ \emph {et~al.}(2016)\citenamefont {Jia},
  \citenamefont {Barth}, \citenamefont {Edwards}, \citenamefont {Mikhailova},\
  and\ \citenamefont {Fisch}}]{Jia2016}%
  \BibitemOpen
  \bibfield  {author} {\bibinfo {author} {\bibfnamefont {Q.}~\bibnamefont
  {Jia}}, \bibinfo {author} {\bibfnamefont {I.}~\bibnamefont {Barth}}, \bibinfo
  {author} {\bibfnamefont {M.~R.}\ \bibnamefont {Edwards}}, \bibinfo {author}
  {\bibfnamefont {J.~M.}\ \bibnamefont {Mikhailova}}, \ and\ \bibinfo {author}
  {\bibfnamefont {N.~J.}\ \bibnamefont {Fisch}},\ }\bibfield  {title} {\bibinfo
  {title} {{Distinguishing Raman from strongly coupled Brillouin amplification
  for short pulses}},\ }\href@noop {} {\bibfield  {journal} {\bibinfo
  {journal} {Phys. Plasmas}\ }\textbf {\bibinfo {volume} {23}},\ \bibinfo
  {pages} {053118} (\bibinfo {year} {2016})}\BibitemShut {NoStop}%
\bibitem [{\citenamefont {Humphrey}\ \emph {et~al.}(2013)\citenamefont
  {Humphrey}, \citenamefont {Trines}, \citenamefont {Fiuza}, \citenamefont
  {Speirs}, \citenamefont {Norreys}, \citenamefont {Cairns}, \citenamefont
  {Silva},\ and\ \citenamefont {Bingham}}]{Humphrey2013effect}%
  \BibitemOpen
  \bibfield  {author} {\bibinfo {author} {\bibfnamefont {K.}~\bibnamefont
  {Humphrey}}, \bibinfo {author} {\bibfnamefont {R.}~\bibnamefont {Trines}},
  \bibinfo {author} {\bibfnamefont {F.}~\bibnamefont {Fiuza}}, \bibinfo
  {author} {\bibfnamefont {D.}~\bibnamefont {Speirs}}, \bibinfo {author}
  {\bibfnamefont {P.}~\bibnamefont {Norreys}}, \bibinfo {author} {\bibfnamefont
  {R.}~\bibnamefont {Cairns}}, \bibinfo {author} {\bibfnamefont
  {L.}~\bibnamefont {Silva}}, \ and\ \bibinfo {author} {\bibfnamefont
  {R.}~\bibnamefont {Bingham}},\ }\bibfield  {title} {\bibinfo {title} {{Effect
  of collisions on amplification of laser beams by Brillouin scattering in
  plasmas}},\ }\href@noop {} {\bibfield  {journal} {\bibinfo  {journal} {Phys.
  Plasmas}\ }\textbf {\bibinfo {volume} {20}},\ \bibinfo {pages} {102114}
  (\bibinfo {year} {2013})}\BibitemShut {NoStop}%
\bibitem [{\citenamefont {Balakin}\ \emph {et~al.}(2011)\citenamefont
  {Balakin}, \citenamefont {Fisch}, \citenamefont {Fraiman}, \citenamefont
  {Malkin},\ and\ \citenamefont {Toroker}}]{Balakin2011numerical}%
  \BibitemOpen
  \bibfield  {author} {\bibinfo {author} {\bibfnamefont {A.}~\bibnamefont
  {Balakin}}, \bibinfo {author} {\bibfnamefont {N.}~\bibnamefont {Fisch}},
  \bibinfo {author} {\bibfnamefont {G.}~\bibnamefont {Fraiman}}, \bibinfo
  {author} {\bibfnamefont {V.}~\bibnamefont {Malkin}}, \ and\ \bibinfo {author}
  {\bibfnamefont {Z.}~\bibnamefont {Toroker}},\ }\bibfield  {title} {\bibinfo
  {title} {{Numerical modeling of quasitransient backward Raman amplification
  of laser pulses in moderately undercritical plasmas with multicharged
  ions}},\ }\href@noop {} {\bibfield  {journal} {\bibinfo  {journal} {Phys.
  Plasmas}\ }\textbf {\bibinfo {volume} {18}},\ \bibinfo {pages} {102311}
  (\bibinfo {year} {2011})}\BibitemShut {NoStop}%
\bibitem [{\citenamefont {Nicholson}(1983)}]{Nicholson1983}%
  \BibitemOpen
  \bibfield  {author} {\bibinfo {author} {\bibfnamefont {D.~R.}\ \bibnamefont
  {Nicholson}},\ }\href@noop {} {\emph {\bibinfo {title} {Introduction to
  Plasma Theory}}}\ (\bibinfo  {publisher} {John Wiley \& Sons},\ \bibinfo
  {year} {1983})\BibitemShut {NoStop}%
\bibitem [{\citenamefont {Bobroff}\ and\ \citenamefont
  {Haus}(1967)}]{Bobroff1967impulse}%
  \BibitemOpen
  \bibfield  {author} {\bibinfo {author} {\bibfnamefont {D.}~\bibnamefont
  {Bobroff}}\ and\ \bibinfo {author} {\bibfnamefont {H.}~\bibnamefont {Haus}},\
  }\bibfield  {title} {\bibinfo {title} {Impulse response of active coupled
  wave systems},\ }\href@noop {} {\bibfield  {journal} {\bibinfo  {journal} {J.
  Appl, Phys.}\ }\textbf {\bibinfo {volume} {38}},\ \bibinfo {pages} {390}
  (\bibinfo {year} {1967})}\BibitemShut {NoStop}%
\bibitem [{\citenamefont {Ginzburg}(1970)}]{Ginzburg1970}%
  \BibitemOpen
  \bibfield  {author} {\bibinfo {author} {\bibfnamefont {V.~L.}\ \bibnamefont
  {Ginzburg}},\ }\href@noop {} {\emph {\bibinfo {title} {The Propagation of
  Electromagnetic Waves in Plasmas}}}\ (\bibinfo  {publisher} {Pergamon
  Press},\ \bibinfo {year} {1970})\BibitemShut {NoStop}%
\bibitem [{\citenamefont {Epperlein}\ \emph {et~al.}(1992)\citenamefont
  {Epperlein}, \citenamefont {Short},\ and\ \citenamefont
  {Simon}}]{Epperlein1992Damping}%
  \BibitemOpen
  \bibfield  {author} {\bibinfo {author} {\bibfnamefont {E.~M.}\ \bibnamefont
  {Epperlein}}, \bibinfo {author} {\bibfnamefont {R.~W.}\ \bibnamefont
  {Short}}, \ and\ \bibinfo {author} {\bibfnamefont {A.}~\bibnamefont
  {Simon}},\ }\bibfield  {title} {\bibinfo {title} {Damping of ion-acoustic
  waves in the presence of electron-ion collisions},\ }\href@noop {} {\bibfield
   {journal} {\bibinfo  {journal} {Phys. Rev. Lett.}\ }\textbf {\bibinfo
  {volume} {69}},\ \bibinfo {pages} {1765} (\bibinfo {year}
  {1992})}\BibitemShut {NoStop}%
\bibitem [{\citenamefont {Kruer}(2003)}]{Kruer2003}%
  \BibitemOpen
  \bibfield  {author} {\bibinfo {author} {\bibfnamefont {W.~L.}\ \bibnamefont
  {Kruer}},\ }\href@noop {} {\emph {\bibinfo {title} {The Physics of Laser
  Plasma Interactions}}}\ (\bibinfo  {publisher} {Westview Press},\ \bibinfo
  {year} {2003})\BibitemShut {NoStop}%
\bibitem [{\citenamefont {Clark}(2003)}]{ClarkThesis}%
  \BibitemOpen
  \bibfield  {author} {\bibinfo {author} {\bibfnamefont {D.~S.}\ \bibnamefont
  {Clark}},\ }\emph {\bibinfo {title} {{Investigations of Raman Amplification
  in Preformed and Ionizing Plasmas}}},\ \href@noop {} {Ph.D. thesis},\
  \bibinfo  {school} {Princeton University} (\bibinfo {year}
  {2003})\BibitemShut {NoStop}%
\bibitem [{\citenamefont {Clark}\ and\ \citenamefont
  {Fisch}(2003{\natexlab{a}})}]{Clark2003operating}%
  \BibitemOpen
  \bibfield  {author} {\bibinfo {author} {\bibfnamefont {D.~S.}\ \bibnamefont
  {Clark}}\ and\ \bibinfo {author} {\bibfnamefont {N.~J.}\ \bibnamefont
  {Fisch}},\ }\bibfield  {title} {\bibinfo {title} {{Operating regime for a
  backward Raman laser amplifier in preformed plasma}},\ }\href@noop {}
  {\bibfield  {journal} {\bibinfo  {journal} {Phys. Plasmas}\ }\textbf
  {\bibinfo {volume} {10}},\ \bibinfo {pages} {3363} (\bibinfo {year}
  {2003}{\natexlab{a}})}\BibitemShut {NoStop}%
\bibitem [{\citenamefont {Clark}\ and\ \citenamefont
  {Fisch}(2003{\natexlab{b}})}]{Clark2003particle}%
  \BibitemOpen
  \bibfield  {author} {\bibinfo {author} {\bibfnamefont {D.~S.}\ \bibnamefont
  {Clark}}\ and\ \bibinfo {author} {\bibfnamefont {N.~J.}\ \bibnamefont
  {Fisch}},\ }\bibfield  {title} {\bibinfo {title} {{Particle-in-cell
  simulations of Raman laser amplification in preformed plasmas}},\ }\href@noop
  {} {\bibfield  {journal} {\bibinfo  {journal} {Phys. Plasmas}\ }\textbf
  {\bibinfo {volume} {10}},\ \bibinfo {pages} {4848} (\bibinfo {year}
  {2003}{\natexlab{b}})}\BibitemShut {NoStop}%
\bibitem [{\citenamefont {Arber}\ \emph {et~al.}(2015)\citenamefont {Arber},
  \citenamefont {Bennett}, \citenamefont {Brady}, \citenamefont
  {Lawrence-Douglas}, \citenamefont {Ramsay}, \citenamefont {Sircombe},
  \citenamefont {Gillies}, \citenamefont {Evans}, \citenamefont {Schmitz},
  \citenamefont {Bell},\ and\ \citenamefont {Ridgers}}]{Arber2015contemporary}%
  \BibitemOpen
  \bibfield  {author} {\bibinfo {author} {\bibfnamefont {T.~D.}\ \bibnamefont
  {Arber}}, \bibinfo {author} {\bibfnamefont {K.}~\bibnamefont {Bennett}},
  \bibinfo {author} {\bibfnamefont {C.~S.}\ \bibnamefont {Brady}}, \bibinfo
  {author} {\bibfnamefont {A.}~\bibnamefont {Lawrence-Douglas}}, \bibinfo
  {author} {\bibfnamefont {M.~G.}\ \bibnamefont {Ramsay}}, \bibinfo {author}
  {\bibfnamefont {N.~J.}\ \bibnamefont {Sircombe}}, \bibinfo {author}
  {\bibfnamefont {P.}~\bibnamefont {Gillies}}, \bibinfo {author} {\bibfnamefont
  {R.~G.}\ \bibnamefont {Evans}}, \bibinfo {author} {\bibfnamefont
  {H.}~\bibnamefont {Schmitz}}, \bibinfo {author} {\bibfnamefont {A.~R.}\
  \bibnamefont {Bell}}, \ and\ \bibinfo {author} {\bibfnamefont {C.~P.}\
  \bibnamefont {Ridgers}},\ }\bibfield  {title} {\bibinfo {title} {Contemporary
  particle-in-cell approach to laser-plasma modelling},\ }\href@noop {}
  {\bibfield  {journal} {\bibinfo  {journal} {Plasma Phys. Contr. F.}\ }\textbf
  {\bibinfo {volume} {57}},\ \bibinfo {pages} {113001} (\bibinfo {year}
  {2015})}\BibitemShut {NoStop}%
\bibitem [{\citenamefont {Weir}\ \emph {et~al.}(1996)\citenamefont {Weir},
  \citenamefont {Mitchell},\ and\ \citenamefont
  {Nellis}}]{Weir1996metallization}%
  \BibitemOpen
  \bibfield  {author} {\bibinfo {author} {\bibfnamefont {S.~T.}\ \bibnamefont
  {Weir}}, \bibinfo {author} {\bibfnamefont {A.~C.}\ \bibnamefont {Mitchell}},
  \ and\ \bibinfo {author} {\bibfnamefont {W.~J.}\ \bibnamefont {Nellis}},\
  }\bibfield  {title} {\bibinfo {title} {Metallization of fluid molecular
  hydrogen at 140 {GPa (1.4 Mbar)}},\ }\href@noop {} {\bibfield  {journal}
  {\bibinfo  {journal} {Phys. Rev. Lett.}\ }\textbf {\bibinfo {volume} {76}},\
  \bibinfo {pages} {1860} (\bibinfo {year} {1996})}\BibitemShut {NoStop}%
\bibitem [{\citenamefont {Balakin}\ \emph {et~al.}(2003)\citenamefont
  {Balakin}, \citenamefont {Fraiman}, \citenamefont {Fisch},\ and\
  \citenamefont {Malkin}}]{Balakin2003noise}%
  \BibitemOpen
  \bibfield  {author} {\bibinfo {author} {\bibfnamefont {A.~A.}\ \bibnamefont
  {Balakin}}, \bibinfo {author} {\bibfnamefont {G.~M.}\ \bibnamefont
  {Fraiman}}, \bibinfo {author} {\bibfnamefont {N.~J.}\ \bibnamefont {Fisch}},
  \ and\ \bibinfo {author} {\bibfnamefont {V.~M.}\ \bibnamefont {Malkin}},\
  }\bibfield  {title} {\bibinfo {title} {Noise suppression and enhanced
  focusability in plasma {Raman} amplifier with multi-frequency pump},\ }\href
  {\doibase 10.1063/1.1621002} {\bibfield  {journal} {\bibinfo  {journal}
  {Phys. Plasmas}\ }\textbf {\bibinfo {volume} {10}},\ \bibinfo {pages} {4856}
  (\bibinfo {year} {2003})}\BibitemShut {NoStop}%
\bibitem [{\citenamefont {Solodov}\ \emph {et~al.}(2003)\citenamefont
  {Solodov}, \citenamefont {Malkin},\ and\ \citenamefont
  {Fisch}}]{Solodov2003}%
  \BibitemOpen
  \bibfield  {author} {\bibinfo {author} {\bibfnamefont {A.}~\bibnamefont
  {Solodov}}, \bibinfo {author} {\bibfnamefont {V.}~\bibnamefont {Malkin}}, \
  and\ \bibinfo {author} {\bibfnamefont {N.}~\bibnamefont {Fisch}},\ }\bibfield
   {title} {\bibinfo {title} {Random density inhomogeneities and focusability
  of the output pulses for plasma-based powerful backward {Raman} amplifiers},\
  }\href@noop {} {\bibfield  {journal} {\bibinfo  {journal} {Phys. Plasmas}\
  }\textbf {\bibinfo {volume} {10}},\ \bibinfo {pages} {2540} (\bibinfo {year}
  {2003})}\BibitemShut {NoStop}%
\bibitem [{\citenamefont {Fraiman}\ \emph {et~al.}(2002)\citenamefont
  {Fraiman}, \citenamefont {Yampolsky}, \citenamefont {Malkin},\ and\
  \citenamefont {Fisch}}]{Fraiman2002}%
  \BibitemOpen
  \bibfield  {author} {\bibinfo {author} {\bibfnamefont {G.~M.}\ \bibnamefont
  {Fraiman}}, \bibinfo {author} {\bibfnamefont {N.~A.}\ \bibnamefont
  {Yampolsky}}, \bibinfo {author} {\bibfnamefont {V.~M.}\ \bibnamefont
  {Malkin}}, \ and\ \bibinfo {author} {\bibfnamefont {N.~J.}\ \bibnamefont
  {Fisch}},\ }\bibfield  {title} {\bibinfo {title} {{Robustness of laser phase
  fronts in backward Raman amplifiers}},\ }\href@noop {} {\bibfield  {journal}
  {\bibinfo  {journal} {Phys. Plasmas}\ }\textbf {\bibinfo {volume} {9}},\
  \bibinfo {pages} {3617} (\bibinfo {year} {2002})}\BibitemShut {NoStop}%
\bibitem [{\citenamefont {Kirkwood}\ \emph {et~al.}(2011)\citenamefont
  {Kirkwood}, \citenamefont {Ping}, \citenamefont {Wilks}, \citenamefont
  {Meezan}, \citenamefont {Michel}, \citenamefont {Williams}, \citenamefont
  {Clark}, \citenamefont {Suter}, \citenamefont {Landen}, \citenamefont {Fisch}
  \emph {et~al.}}]{Kirkwood2011observation}%
  \BibitemOpen
  \bibfield  {author} {\bibinfo {author} {\bibfnamefont {R.}~\bibnamefont
  {Kirkwood}}, \bibinfo {author} {\bibfnamefont {Y.}~\bibnamefont {Ping}},
  \bibinfo {author} {\bibfnamefont {S.}~\bibnamefont {Wilks}}, \bibinfo
  {author} {\bibfnamefont {N.}~\bibnamefont {Meezan}}, \bibinfo {author}
  {\bibfnamefont {P.}~\bibnamefont {Michel}}, \bibinfo {author} {\bibfnamefont
  {E.}~\bibnamefont {Williams}}, \bibinfo {author} {\bibfnamefont
  {D.}~\bibnamefont {Clark}}, \bibinfo {author} {\bibfnamefont
  {L.}~\bibnamefont {Suter}}, \bibinfo {author} {\bibfnamefont
  {O.}~\bibnamefont {Landen}}, \bibinfo {author} {\bibfnamefont
  {N.}~\bibnamefont {Fisch}},  \emph {et~al.},\ }\bibfield  {title} {\bibinfo
  {title} {Observation of amplification of light by {Langmuir} waves and its
  saturation on the electron kinetic timescale},\ }\href@noop {} {\bibfield
  {journal} {\bibinfo  {journal} {J. Plasma Phys.}\ }\textbf {\bibinfo {volume}
  {77}},\ \bibinfo {pages} {521} (\bibinfo {year} {2011})}\BibitemShut
  {NoStop}%
\bibitem [{\citenamefont {Geloni}\ \emph {et~al.}(2010)\citenamefont {Geloni},
  \citenamefont {Saldin}, \citenamefont {Samoylova}, \citenamefont
  {Schneidmiller}, \citenamefont {Sinn}, \citenamefont {Tschentscher},\ and\
  \citenamefont {Yurkov}}]{Geloni2010coherence}%
  \BibitemOpen
  \bibfield  {author} {\bibinfo {author} {\bibfnamefont {G.}~\bibnamefont
  {Geloni}}, \bibinfo {author} {\bibfnamefont {E.}~\bibnamefont {Saldin}},
  \bibinfo {author} {\bibfnamefont {L.}~\bibnamefont {Samoylova}}, \bibinfo
  {author} {\bibfnamefont {E.}~\bibnamefont {Schneidmiller}}, \bibinfo {author}
  {\bibfnamefont {H.}~\bibnamefont {Sinn}}, \bibinfo {author} {\bibfnamefont
  {T.}~\bibnamefont {Tschentscher}}, \ and\ \bibinfo {author} {\bibfnamefont
  {M.}~\bibnamefont {Yurkov}},\ }\bibfield  {title} {\bibinfo {title}
  {Coherence properties of the {European XFEL}},\ }\href@noop {} {\bibfield
  {journal} {\bibinfo  {journal} {New J. Phys.}\ }\textbf {\bibinfo {volume}
  {12}},\ \bibinfo {pages} {035021} (\bibinfo {year} {2010})}\BibitemShut
  {NoStop}%
\bibitem [{\citenamefont {Vartanyants}\ \emph {et~al.}(2011)\citenamefont
  {Vartanyants}, \citenamefont {Singer}, \citenamefont {Mancuso}, \citenamefont
  {Yefanov}, \citenamefont {Sakdinawat}, \citenamefont {Liu}, \citenamefont
  {Bang}, \citenamefont {Williams}, \citenamefont {Cadenazzi}, \citenamefont
  {Abbey} \emph {et~al.}}]{Vartanyants2011coherence}%
  \BibitemOpen
  \bibfield  {author} {\bibinfo {author} {\bibfnamefont {I.}~\bibnamefont
  {Vartanyants}}, \bibinfo {author} {\bibfnamefont {A.}~\bibnamefont {Singer}},
  \bibinfo {author} {\bibfnamefont {A.}~\bibnamefont {Mancuso}}, \bibinfo
  {author} {\bibfnamefont {O.}~\bibnamefont {Yefanov}}, \bibinfo {author}
  {\bibfnamefont {A.}~\bibnamefont {Sakdinawat}}, \bibinfo {author}
  {\bibfnamefont {Y.}~\bibnamefont {Liu}}, \bibinfo {author} {\bibfnamefont
  {E.}~\bibnamefont {Bang}}, \bibinfo {author} {\bibfnamefont {G.~J.}\
  \bibnamefont {Williams}}, \bibinfo {author} {\bibfnamefont {G.}~\bibnamefont
  {Cadenazzi}}, \bibinfo {author} {\bibfnamefont {B.}~\bibnamefont {Abbey}},
  \emph {et~al.},\ }\bibfield  {title} {\bibinfo {title} {Coherence properties
  of individual femtosecond pulses of an x-ray free-electron laser},\
  }\href@noop {} {\bibfield  {journal} {\bibinfo  {journal} {Phys. Rev. Lett.}\
  }\textbf {\bibinfo {volume} {107}},\ \bibinfo {pages} {144801} (\bibinfo
  {year} {2011})}\BibitemShut {NoStop}%
\end{thebibliography}
\end{document}